\documentclass{LMCS}

\usepackage{amssymb}
\usepackage{amsmath}
\usepackage{bussproofs}
\usepackage{prooftree}
\usepackage{txfonts}
\usepackage{diagrams}
\usepackage{enumerate,hyperref}

\usepackage{graphicx}

\newarrow{Map}{vee}---{>}
\newarrow{Rel}{}-{+}-{>}
\newarrow{Small}{}{}{}-{>}

\newcommand{\ILLomega}{{\sf ILL}^\omega}
\newcommand{\ILLomegab}{{\sf ILL}^\omega_b}
\newcommand{\ILomega}{{\sf IL}^\omega}
\newcommand{\ILLomegar}{{\sf ILL}^\omega_{r}}
\newcommand{\DC}{{\sf DC}}
\newcommand{\C}{{\sf C}}

\newcommand{\itype}{i}
\newcommand{\btype}{b}

\newcommand{\cwedge}{\otimes}
\newcommand{\awedge}{\,\&\,}
\newcommand{\bang}[1]{! #1}
\newcommand{\lto}{\multimap}
\newcommand{\lequiv}{\multimapboth}
\newcommand{\pcond}[3]{#2 \; \Diamond_{#1} \, #3}
\newcommand{\true}{\textsc{T}}
\newcommand{\false}{\textsc{F}}
\newcommand{\be}{=^b}
\newcommand{\aeq}{\rotatebox[origin = c]{180}{{\sf \AE}}}
\newcommand{\sq}[2]{{\aeq}^{#1}_{\!#2}}

\newcommand{\Pdis}{\textup{P}_\oplus}
\newcommand{\Pexi}{\textup{P}_\exists}

\newcommand{\ruleid}{\text{id}}
\newcommand{\rulecut}{\text{cut}}
\newcommand{\ruleper}{\text{per}}
\newcommand{\rulecon}{\text{con}}
\newcommand{\rulewkn}{\text{wkn}}

\newcommand{\mr}{\;{\sf mr}\;}
\newcommand{\uInter}[3]{|#1|^{#2}_{#3}}
\newcommand{\dInter}[3]{#1_{D}(#2;#3)}
\newcommand{\dnInter}[3]{#1_{dn}(#2;#3)}
\newcommand{\lTrans}[1]{#1^{*}}
\newcommand{\bTrans}[1]{#1^{\circ}}
\newcommand{\pTrans}[1]{#1^{+}}

\newcommand{\lAC}{{\sf AC}_l}
\newcommand{\lMP}{{\sf MP}_l}
\newcommand{\lIP}{{\sf IP}_l}
\newcommand{\EP}{{\sf EP}}

\newcommand{\Aat}{A_{{\sf at}}}
\newcommand{\Aqf}{A_{{\sf qf}}}
\newcommand{\Bqf}{B_{{\sf qf}}}

\newcommand{\ubq}[3]{\forall #1 \!\sqsubset\! #2 \, #3}
\newcommand{\singleton}[1]{\eta #1}
\newcommand{\join}[2]{#1 \otimes #2}
\newcommand{\comp}[2]{#1 \circ #2}

\newcommand{\FV}[1]{{\sf FV}(#1)}
\newcommand{\pdefin}{:\equiv}
\newcommand{\eqleft}[1]{\begin{itemize} \item[] $#1$ \end{itemize}}
\newcommand{\pvec}[1]{\boldsymbol{#1}}
\newcommand{\proves}{\vdash}

\def\doi{7 (1:9) 2011}
\lmcsheading%
{\doi}
{1--22}
{}
{}
{Mar.~10, 2010}
{Mar.~24, 2011}
{}   

\begin{document}

\title{Functional Interpretations of Intuitionistic Linear Logic}
\author[G.~Ferreira]{Gilda Ferreira\rsuper a}
\address{{\lsuper a}Departamento de Matem\'atica, Faculdade de Ci\^{e}ncias da Universidade de Lisboa}
\email{gildafer@cii.fc.ul.pt}
\thanks{{\lsuper a}The first author would like to thank Funda\c{c}\~{a}o para a Ci\^{e}ncia e a Tecnologia (grant SFRH/BPD/34527/2006 and project {\sc PTDC/MAT/104716/2008}) and Centro de Matem\'{a}tica e Aplica\c{c}\~{o}es Fundamentais.}

\author[P.~Oliva]{Paulo Oliva\rsuper b}
\address{{\lsuper b}Queen Mary University of London, School of Electronic
  Engineering and Computer Science}
\email{paulo.oliva@eecs.qmul.ac.uk}
\thanks{{\lsuper b}The second author gratefully acknowledges support of the Royal Society (grant number 516002.K501/RH/kk).}

\begin{abstract} We present three different functional interpretations of \emph{intuitionistic linear logic} and show how these correspond to well-known functional interpretations of intuitionistic logic via embeddings of $\ILomega$ into $\ILLomega$. The main difference from previous work of the second author is that in intuitionistic linear logic (as opposed to classical linear logic) the interpretations of $\bang A$ are simpler and simultaneous quantifiers are no longer needed for the characterisation of the interpretations. We then compare our approach in developing these three proof interpretations with the one of de Paiva around the Dialectica category model of linear logic.
\end{abstract}

\maketitle

\section{Introduction}

This paper presents a family of functional interpretations of \emph{intuitionistic} linear logic. First, we present a single functional interpretation of pure (i.e., the exponential-free fragment of) intuitionistic linear logic. This is followed by a parametrised interpretation of the exponential $\bang A$. Finally, three possible instances of the parameter are considered and shown to correspond to three well-known functional interpretation of intuitionistic logic.

The second author \cite{Oliva(2007A),Oliva(2007),Oliva(2008),Oliva(2009A)} has recently shown how different functional interpretations of intuitionistic logic can be factored into a uniform family of interpretations of classical linear logic combined with Girard's standard embedding $\lTrans{(\cdot)}$ of intuitionistic logic into linear logic (see also \cite{GO(2010)}). In the symmetric context of \emph{classical linear logic} each formula $A$ is associated with a simultaneous one-move two-player game $\uInter{A}{\pvec x}{\pvec y}$. Intuitively, the two players, say Eloise and Abelard, must pick their moves $\pvec x$ and $\pvec y$ simultaneously and Eloise wins if and only if $\uInter{A}{\pvec x}{\pvec y}$ holds. The symmetric nature of the game implies that (proof-theoretically) the formula $A$ is interpreted as the formula
\[ \sq{\pvec x}{\pvec y} \uInter{A}{\pvec x}{\pvec y} \]
where $\sq{\pvec x}{\pvec y} A$ is a simple form of branching quantifier -- termed \emph{simultaneous quantifier} in \cite{Oliva(2007)}. Following this game-theoretic reading, the different interpretations of the modality $\bang A$ are all of the following form: First, it (always) turns a symmetric game into an asymmetric one, where Eloise plays first, giving Abelard the advantage of playing second. In the symmetric context, this asymmetric game can be modelled by allowing Abelard to play a function $\pvec f$ which calculates his move from a given Eloise move $\pvec x$. Secondly, the game $\bang A$ gives a further (non-canonical) advantage to Abelard, by allowing him to play a \emph{set of moves}, rather than a single move. The idea is the following: Abelard wins the game $\bang A$ if there is a move $\pvec y \in \pvec f \pvec x$ that is winning with respect to Eloise's move $\pvec x$, i.e. $\neg \uInter{A}{\pvec x}{\pvec y}$. Formally
\[ \uInter{\bang A}{\pvec x}{\pvec f} \; \equiv \; \forall \pvec y \!\in\! \pvec f \pvec x \, \uInter{A}{\pvec x}{\pvec y}. \]
Therefore, the game $\bang A$ always introduces a break of symmetry, but it leaves open \emph{what kind of sets} Abelard is allowed to play. What the second author has shown is that if only singleton sets are allowed the resulting interpretation corresponds to G\"odel's Dialectica interpretation \cite{Avigad(98),Goedel(58),Oliva(2008)}; if finite sets are allowed then it corresponds to the Diller-Nahm variant of the Dialectica interpretation \cite{Diller(74),Oliva(2009A)}; and if these sets are actually the whole set of moves then it corresponds to Kreisel's modified realizability interpretation \cite{Kreisel(59),Oliva(2007)}.

In the present paper we show that in the context of \emph{intuitionistic linear
logic} every formula can be interpreted as a game where Eloise plays
first and Abelard plays second, the branching quantifiers being no
longer needed. In other words, Abelard's advantage of playing
second, which was limited to the game $\bang A$ in classical linear
logic, is ubiquitous in intuitionistic linear logic. In this way,
the game-theoretic interpretation of the modality $\bang A$ is
simply to lift the moves of Abelard from a single move to a set of
moves. Formally,
\[ \uInter{\bang A}{\pvec x}{\pvec a} \;\equiv\; \forall \pvec y \!\in\! \pvec a \, \uInter{A}{\pvec x}{\pvec y}. \]
Therefore, by working in the context of $\ILLomega$, we can fully separate the canonical part of the interpretation (pure intuitionistic linear logic), where all interpretations coincide, and the non-canonical part where each choice of ``sets of moves" gives rise to a different functional interpretation.

As we shall see, the functional interpretation of \emph{pure intuitionistic linear logic} coincides with G\"odel's Dialectica interpretation of \emph{intuitionistic logic}, reading $\lto, \cwedge$ and $\oplus$ as $\to, \wedge$ and $\vee$, respectively. This is so because the Dialectica interpretation identifies the games $A$ and $\bang A$. The connection between G\"odel's Dialectica interpretation and intuitionistic linear logic was first studied by de Paiva \cite{dePaiva(1989A)}. One can view our work here as a proof-theoretic reading of de Paiva's category-theoretic work, together with an extension linking the ``Dialectica" interpretation of intuitionistic linear logic also with Kreisel's modified realizability (see also Biering's recent work \cite{Biering(2008)}).

The paper is organised as follows: In Section \ref{sec:basic} we present the basic interpretation of pure intuitionistic linear logic. In the same section we outline which principles are needed for the characterisation of the interpretation (Subsection \ref{sec:characterisation}). Section \ref{sec:modal} describes three different interpretations of the modality $\bang A$. This is followed (Section \ref{sec:relation}) by a description of how these choices correspond to three well-known functional interpretations of intuitionistic logic: Kreisel's modified realizability, Diller-Nahm interpretation and G\"{o}del's Dialectica interpretation. Finally, in Section \ref{categorical} we compare our approach (based on finite types) with that of de Paiva (based on cartesian closed categories).

\subsection{Intuitionistic Linear Logic}

Intuitionistic linear logic can be viewed as a fragment of Girard's linear logic \cite{Girard(87B)} which is sufficient for embedding intuitionistic
logic into the linear context. We will make use of the formulation of intuitionistic linear logic
shown in Tables \ref{ill-connectives} and \ref{ill-quantifiers} with the usual side conditions in the rules $\forall\textup{R}$ and $\exists \textup{L}$. Our system is denoted by $\ILLomega$
since we work in the language of all finite types.

\begin{table}[t]
\[
\begin{array}{|rccc|}
\hline
& & & \\
\hspace{15mm} & \begin{prooftree} \justifies A \proves A \using (\ruleid) \end{prooftree}
& \hspace{5mm} &
\begin{prooftree} \justifies \Gamma, 0 \proves A  \end{prooftree} \quad \quad \\[5mm]
& \begin{prooftree} \Gamma \proves A \quad \Delta, A \proves B \justifies \Gamma, \Delta \proves B \using (\rulecut) \end{prooftree}
& &
\begin{prooftree} \Gamma \proves A \justifies \pi\{\Gamma\} \proves A \using (\ruleper) \end{prooftree} \\[5mm]
\hline
& & & \\
& \begin{prooftree} \Gamma \proves A \quad \Delta \proves B \justifies \Gamma,\Delta \proves A \cwedge B \using (\cwedge\textup{R}) \end{prooftree}
& &
\begin{prooftree} \Gamma, A, B \proves C \justifies \Gamma, A \cwedge B \proves C \using (\cwedge\textup{L}) \end{prooftree} \\[5mm]
& \begin{prooftree} \Gamma, A \proves B \justifies \Gamma \proves A \lto B \using (\lto\!\textup{R}) \end{prooftree}
& &
\begin{prooftree} \Gamma \proves A \quad \Delta, B \proves C \justifies \Gamma, \Delta, A \lto B \proves C \using (\lto\!\textup{L}) \end{prooftree} \\[5mm]
\hline
& & & \\
\multicolumn{4}{|l|}{
\quad
\begin{array}{crcrc}
\begin{prooftree} \Gamma \proves A \quad \Gamma \proves B\justifies  \Gamma \proves A \awedge B \using (\&\textup{R}) \end{prooftree}
& \quad &
\begin{prooftree} \Gamma, A \proves B\justifies  \Gamma, A\awedge  C \proves B \using (\&\textup{L}) \end{prooftree}
& \quad &
\begin{prooftree} \Gamma, A \proves B \justifies \Gamma, C \awedge A \proves B \using (\&\textup{L}) \end{prooftree} \\[5mm]
\begin{prooftree} \Gamma \proves A\justifies  \Gamma\proves A \oplus B \using (\oplus\textup{R}) \end{prooftree}
& &
\begin{prooftree} \Gamma \proves B\justifies  \Gamma\proves A \oplus B \using (\oplus\textup{R}) \end{prooftree}
& &
\begin{prooftree} \Gamma, A \proves C \quad \Gamma, B \proves C\justifies  \Gamma, A \oplus B \proves C \using (\oplus\textup{L})  \quad\end{prooftree} \\[5mm]
\end{array}
} \\[5mm]
\hline
\end{array}
\]
\caption{Intuitionistic Linear Logic (connectives)} \label{ill-connectives}
\end{table}

The finite types are inductively defined in the usual way: $\itype$ is a finite type and if
$\rho$ and $\sigma$ are finite types then $\rho\to \sigma$ is a
finite type. The terms of $\ILLomega$ are: the constants (including one of type $\itype$ to ensure that
all types are inhabited by a closed term and the
typed combinators $\Pi^{\sigma\to\tau\to\sigma}$ and
$\Sigma^{(\rho\to\sigma\to\tau)\to(\rho\to\sigma)\to\rho\to\tau}$),
infinitely many variables $x^{\rho}$ of each finite type $\rho$, and if
$t^{\sigma\to\tau}$ and $s^{\sigma}$ are terms then the application $t s$ is a term
of type $\tau$. We assume a neutral treatment of equality in the system $\ILLomega$ (cf. \cite{Troelstra(73)}), i.e. the combinators are axiomatised as 
\[ A[\Pi xy/w]\lequiv  A[x/w] \quad \quad \mbox{and} \quad \quad  A[\Sigma xyz/w] \lequiv  A[xz(yz)/w], \]
where $A \lequiv B$ is an abbreviation from $(A \lto B) \, \& \, (B \lto A)$. By \emph{combinatorial completeness}, we know that we can associate with
each term $t^{\sigma}$ and variable $x^{\tau}$ a term $\lambda x.t$
of type $\tau\to\sigma$ also satisfying $ A[(\lambda x.t)(s)/w]\lequiv
A[t[s/x]/w]$. 

The atomic formulas of $\ILLomega$ are denoted by $\Aat$ (the linear logic constant 0 is an
atomic formula) and if $A$ and $B$ are formulas, then $A\cwedge B$,
$A \awedge B$, $A\oplus B$, $A\lto B$, $\bang A$, $\forall x A(x)$ and
$\exists x A(x)$ are also formulas.

In this paper we will also work with a subsystem of $\ILLomega$, dubbed $\ILLomegar$, where the following restriction is
assumed on the $\&\textup{R}$-rule: The context $\Gamma$ must consist entirely of formulas of the kind $\bang A$. In Section \ref{sec:relation} we will see why we need this technical restriction. Nevertheless, note that both systems
$\ILLomega$ and $\ILLomegar$ are strong enough to capture intuitionistic
logic $\ILomega$ into the linear context, as made precise in the following proposition.

\begin{prop}[\cite{Girard(87B)}] \label{g-trans} Define two translations of formulas of $\ILomega$ into formulas of $\ILLomegar$ inductively as follows:
\eqleft{
\begin{array}{llll}
\lTrans{\Aat} & \pdefin \Aat
 & \bTrans{\Aat} & \pdefin \;\bang \Aat, \quad \text{ if } \Aat \not\equiv \bot \\[2mm]
\lTrans{\bot} & \pdefin 0
 & \bTrans{\bot} &\pdefin 0 \\[2mm]
\lTrans{(A \wedge B)} & \pdefin \lTrans{A} \awedge \lTrans{B}
 &  \bTrans{(A \wedge B)}     &\pdefin \bTrans{A} \cwedge \bTrans{B} \\[2mm]
\lTrans{(A \vee B)}     &\pdefin \;\bang \lTrans{A} \oplus \;\bang\lTrans{B}
 & \bTrans{(A \vee B)}     &\pdefin  \bTrans{A} \oplus \bTrans{B} \\[2mm]
\lTrans{(A \to B)}     &\pdefin \;\bang \lTrans{A} \lto \lTrans{B} \hspace{15mm}
 & \bTrans{(A \to B)}     &\pdefin \;\bang (\bTrans{A} \lto \bTrans{B}) \\[2mm]
\lTrans{(\forall x A)} &\pdefin \forall x \lTrans{A}
 & \bTrans{(\forall x A)} &\pdefin \;\bang \forall x \bTrans{A} \\[2mm]
\lTrans{(\exists x A)} &\pdefin \exists x \bang \lTrans{A}
 &  \bTrans{(\exists x A)} &\pdefin \exists x \bTrans{A}
\end{array}
}
If $A$ is provable in $\ILomega$ then $\lTrans{A}$ and $\bTrans{A}$
are provable in $\ILLomegar$ (and hence also in $\ILLomega$). Moreover, it is easy to check that $\bTrans{A}
\lequiv \;\bang \lTrans{A}$.
\end{prop}
\proof It is already known that if $\Gamma \proves_{\ILomega} A$ then
$\bang \lTrans{\Gamma} \proves_{\ILLomega} \lTrans{A}$ (see \cite{Girard(87B),Schellinx(1991)}). The result
with $\ILLomega$ replaced by $\ILLomegar$ just requires our
attention in the rule $\&\textup{R}$. The
result for $\bTrans{A}$ follows immediately from the fact that in
$\ILLomegar$ we can prove $\bTrans{A}\lequiv ~ \bang \lTrans{A}$.\qed

\begin{table}[t]
\[
\begin{array}{|rccc|}
\hline
\hspace{15mm} & & & \\
& \begin{prooftree} \Gamma \proves A \justifies \Gamma \proves \forall x^{\rho} A \using (\forall \textup{R}) \end{prooftree}
& &
\begin{prooftree} \Gamma, A[t^{\rho}/x] \proves B \justifies \Gamma, \forall x^{\rho} A \proves B \using (\forall \textup{L}) \end{prooftree} \\[5mm]
& \begin{prooftree} \Gamma \proves A[t^{\rho}/x] \justifies \Gamma \proves \exists x^{\rho} A \using (\exists \textup{R}) \end{prooftree}
& &
\begin{prooftree} \Gamma, A \proves B \justifies \Gamma, \exists x^{\rho} A \proves B \using (\exists \textup{L}) \end{prooftree} \\[5mm]
\hline
& & & \\
\multicolumn{4}{|l|}{
  \quad
  \begin{prooftree} \Gamma, \bang A, \bang A \proves B \justifies \Gamma, \bang A \proves B \using (\rulecon) \end{prooftree}
  \quad \quad\quad
  \begin{prooftree} \Gamma \proves B \justifies \Gamma, \bang A \proves B \using (\rulewkn) \end{prooftree}
  \quad \quad
  \begin{prooftree} \bang \Gamma \proves A \justifies \bang \Gamma \proves \bang A \using (\bang\textup{R}) \end{prooftree}
  \quad \quad
  \begin{prooftree} \Gamma, A \proves B \justifies  \Gamma, \bang A \proves B \using (\bang\textup{L}) \quad \end{prooftree}
} \\
& & & \\
\hline
\end{array}
\]
\caption{Intuitionistic Linear Logic (quantifiers and modality)} \label{ill-quantifiers}
\end{table}

The systems $\ILLomega$ and $\ILLomegar$ will be called \emph{interpreted systems}, to distinguish them from the \emph{verifying system} presented in the next subsection. The interpretations we will discuss map formulas and proofs in the interpreted system into formulas and proofs of the verifying system. In order to obtain a general notion of interpretation, we must work with the simplest (yet relevant) interpreted system possible. When extending the interpretation to more complex systems we must then ensure that the extra axioms and rules are also interpreted, but these might be interpreted by some interpretation but not others. For instance, we chose a neutral treatment of equality in the interpreted system because that is what can be interpreted in general, by all three interpretations considered. If one were to add full extensionally the Dialectica interpretation would no longer work, whereas this would be no problem for the realizability interpretation.\bigskip

\noindent {\bf Notation.} Throughout the paper, boldface letters $\pvec a$, $\pvec b$, $\ldots$ or $\pvec x$, $\pvec y$, $\ldots$ stand for tuples of terms or variables.

\subsection{Verifying system}

As we will show in the next sections, the three presented functional interpretations translate the formula $A \oplus B$ via a sort of flagged disjoint union, i.e. a boolean and a witness for either $A$ or $B$. Therefore, in the verifying system, which we shall denote by
$\ILLomegab$, we assume that the language also contains the booleans $\btype$ as base type. We also assume the existence of two boolean constants true and false ($\true$, $\false$), boolean variables, an equality relation $\be$
between terms of boolean type, and a constant of type $b \to \rho\to\rho\to\rho$ that should be seen as a conditional
$\lambda$-term $z(t,q)$ that equals either $t$ or $q$ depending on whether $z^\btype$ equals true or false.
$\ILLomegab$ is assumed to contain the following axioms for equality:
\begin{enumerate}[(1)]
\item $\bang(x\be x)$
\item $\bang(x\be y)\lto ~\bang(y\be x)$
\item $\bang(x\be y) \cwedge ~\bang (y\be z)\lto ~\bang (x\be z)$
\item $\bang(x\be y)\cwedge A[x/w]\lto A[y/w]$.
\end{enumerate}
We would also like to ensure that true and false are distinct and that
there are no other elements of boolean type
\begin{enumerate}[(1)]
\item[(5)] $\bang(\true\be \false)\lto 0$
\item[(6)] $\bang(z\be \true) \,\oplus\, \bang (z\be \false)$.
\end{enumerate}
The axioms for the conditional $\lambda$-term are as follows
\begin{enumerate}[(1)]
\item[(7)] $A[\true(t,q)/w] \lequiv A[t/w]$ and
$A[\false(t,q)/w]\lequiv A[q/w]$.
\end{enumerate}
For simplicity, we use the following abbreviation:
\eqleft{\pcond{z}{A}{B} \,\pdefin\, (\bang (z\be\true) \lto A) \awedge (\bang (z\be\false) \lto B).}

\begin{lem} \label{useful} The following are derivable in $\ILLomegab$
\begin{enumerate}[\em(i)]
\item
$\begin{prooftree} \proves A[\true] \hspace{.2cm} \proves A[\false]
\justifies
\proves A[z]
\end{prooftree}$ 
\item $\bang(\true \be \true) \lto A\vdash A$ and $\bang(\false \be \false) \lto A\vdash A$
\item $A \vdash ~ \bang(\true\be\false)\lto B$
\item $\pcond{\true}{A}{B}\lequiv A$ and $\pcond{\false}{A}{B}\lequiv B$
\item $\pcond{z}{\bang A}{\bang B}\lequiv \,\bang (\pcond{z}{\bang A}{\bang B}). $
\end{enumerate}
\end{lem}
\proof Assertion (i) can be derived from axioms (4) and (6); (ii) follows
easily from axiom 1.; (iii) can be deduced from axiom (5) and the
forward implications in (iv) follow immediately from item (ii)
and the inverse implications can easily be deduced using (iii).
The forward implication in assertion (v) can be derived using
assertions (i) and (iv), the other implication being trivial.\qed

We stress again that we do not need to worry about which axioms are added to the \emph{verifying system} $\ILLomegab$, as these do not need to be interpreted. For instance, in the verifying system we could even have assumed full extensionality. What we listed above in the description of $\ILLomegab$ is the minimal necessary to verify the basic interpretation of $\ILLomega$ and $\ILLomegar$, to be described in the following section. When extending the basic interpretation to deal with $\bang A$ we will also need to extend the verifying system $\ILLomegab$. The extensions of $\ILLomegab$, however, will depend on the particular interpretation of $\bang A$, and will be introduced in Proposition \ref{banginter} (Section \ref{sec:modal}).

\section{A Basic Interpretation of Pure \texorpdfstring{$\ILLomega$}{ILLomega}}
\label{sec:basic}

In this section we present a basic functional interpretation of pure (without the exponential $\bang A$) intuitionistic linear logic, and
prove its soundness. In the next section we then consider different extensions of this interpretation to full intuitionistic linear
logic.

\begin{defi}[Basic functional interpretation of pure $\ILLomega$] \label{inter} For each formula $A$ of pure $\ILLomega$, let us associate a formula $\uInter{A}{\pvec x}{\pvec y}$ of $\ILLomegab$, with two fresh lists of free-variables $\pvec x$ and $\pvec y$, inductively as follows: For atomic formulas $\Aat$ we let $\uInter{\Aat}{}{} \pdefin \Aat$. Assume the interpretations of $A$ and $B$ have already been defined as $\uInter{A}{\pvec x}{\pvec y}$ and $\uInter{B}{\pvec v}{\pvec w}$, we then define
\eqleft{
\begin{array}{lcl}
\uInter{A \lto B}{\pvec f, \pvec g}{\pvec x, \pvec w} & \pdefin & \uInter{A}{\pvec x}{\pvec f \pvec x \pvec w} \lto \uInter{B}{\pvec g \pvec x}{\pvec w} \\[2mm]
\uInter{A \cwedge B}{\pvec x, \pvec v}{\pvec y, \pvec w} & \pdefin & \uInter{A}{\pvec x}{\pvec y} \cwedge \uInter{B}{\pvec v}{\pvec w} \\[2mm]
\uInter{A \awedge B}{\pvec x, \pvec v}{\pvec y, \pvec w,z} & \pdefin & \pcond{z}{\uInter{A}{\pvec x}{\pvec y}}{\uInter{B}{\pvec v}{\pvec w}} \\[2mm]
\uInter{A \oplus B}{\pvec x, \pvec v,z}{\pvec y, \pvec w} & \pdefin & \pcond{z}{\uInter{A}{\pvec x}{\pvec y}}{\uInter{B}{\pvec v}{\pvec w}} \\[2mm]
\uInter{\exists z A(z)}{\pvec x, z}{\pvec y} & \pdefin & \uInter{A(z)}{\pvec x}{\pvec y} \\[2mm]
\uInter{\forall z A(z)}{\pvec f}{\pvec y, z} & \pdefin &
\uInter{A(z)}{\pvec f z}{\pvec y}.
\end{array}}
\end{defi}

Intuitively, the meaning of $A$ is reduced to the existence of a tuple of
objects $\pvec x$ such that $\forall \pvec y \uInter{A}{\pvec
x}{\pvec y}$. The $\pvec x$'s are called \emph{witnesses} and the
$\pvec y$'s \emph{challenges}. Note that, contrary to the
interpretation of classical linear logic
\cite{Oliva(2007A),Oliva(2009A)}, the functional interpretation of
intuitionistic linear logic is no longer symmetric. In terms of games, the
interpretation above can be seen as associating to each formula $A$
a one-move two-player \emph{sequential} game $\uInter{A}{\pvec x}{\pvec y}$.
In this game, Eloise starts by playing a move $\pvec x$ followed by Abelard
playing a move $\pvec y$. Eloise wins if $\uInter{A}{\pvec x}{\pvec y}$ holds, otherwise Abelard wins.

\begin{thm}[Soundness] \label{soundness} Let $A_0, \ldots, A_n, B$ be formulas of pure $\ILLomega$, with $\pvec z$ as the only free-variables. If
\eqleft{A_0(\pvec z), \ldots, A_n(\pvec z) \proves B(\pvec z)}
is provable in pure $\ILLomega$ then terms $\pvec a_0, \ldots, \pvec
a_n, \pvec b$ can be extracted from this proof such that
\eqleft{\uInter{A_0(\pvec z)}{\pvec x_0}{\pvec a_0}, \ldots,
\uInter{A_n(\pvec z)}{\pvec x_n}{\pvec a_n} \proves \uInter{B(\pvec
z)}{\pvec b}{\pvec w}}
is provable in $\ILLomegab$, with $\FV{\pvec a_i} \subseteq \{\pvec z, \pvec x_0, \ldots, \pvec x_n, \pvec w\}$ and $\FV{\pvec b} \subseteq \{\pvec z, \pvec x_0, \ldots, \pvec x_n\}$.
\end{thm}
\proof By induction on the derivation of $A_0(\pvec z), \ldots, A_n(\pvec z) \proves B(\pvec z)$. The axioms are trivial since the interpretation does not change atomic formulas and every type is inhabited. Note that for the axiom $\Gamma, 0 \proves A$, the interpretation of $0$ is $0$ itself, and so we can take arbitrary terms of the right type. The fact that every type is inhabited is also used in the study of the rules $\&\textup{L}$ and $\oplus\textup{R}$. The permutation rule is immediate. Let us consider a few other cases:\medskip

\noindent\emph{Cut}. By induction hypothesis, assume we already have terms witnessing the two premises  as $\uInter{\Gamma}{\pvec u}{\pvec \gamma} \proves \uInter{A}{\pvec a_0}{\pvec y}$ and $\uInter{\Delta}{\pvec v}{\pvec \delta}, \uInter{A}{\pvec x}{\pvec a_1[\pvec x]} \proves \uInter{B}{\pvec b}{\pvec w}$. We must construct terms that witness the conclusion $\Gamma, \Delta \proves B$. That can be done as follows:
\[
\begin{prooftree}
\[
    \uInter{\Gamma}{\pvec u}{\pvec \gamma} \proves \uInter{A}{\pvec a_0}{\pvec y}
    \justifies
    \uInter{\Gamma}{\pvec u}{\pvec \gamma'} \proves \uInter{A}{\pvec a_0}{\pvec a_1[\pvec a_0]}
    \using {[\frac{\pvec a_1[\pvec a_0]}{\pvec y}]}
\]
\quad
\[
    \uInter{\Delta}{\pvec v}{\pvec \delta}, \uInter{A}{\pvec x}{\pvec a_1[\pvec x]} \proves \uInter{B}{\pvec b}{\pvec w}
    \justifies
    \uInter{\Delta}{\pvec v}{\pvec \delta'}, \uInter{A}{\pvec a_0}{\pvec a_1[\pvec a_0]} \proves \uInter{B}{\pvec b'}{\pvec w}
    \using {[\frac{\pvec a_0}{\pvec x}]}
\]
\justifies
\uInter{\Gamma}{\pvec u}{\pvec \gamma'}, \uInter{\Delta}{\pvec v}{\pvec \delta'}
    \proves \uInter{B}{\pvec b'}{\pvec w}
\using (\rulecut)
\end{prooftree}
\]
where $\pvec \gamma'$ and $\pvec \delta', \pvec b'$ are obtained from $\pvec \gamma$ and $\pvec \delta, \pvec b$ via the substitutions $[\pvec a_1[\pvec a_0]/\pvec y]$ and $[\pvec a_0/\pvec x]$, respectively. \\[2mm]
\emph{Tensor}.
\[
\begin{prooftree}
\[
    \uInter{\Gamma}{\pvec u}{\pvec \gamma} \proves \uInter{A}{\pvec a}{\pvec y}
    \quad
    \uInter{\Delta}{\pvec v}{\pvec \delta} \proves \uInter{B}{\pvec b}{\pvec w}
    \justifies
    \uInter{\Gamma}{\pvec u}{\pvec \gamma}, \uInter{\Delta}{\pvec v}{\pvec \delta}
        \proves \uInter{A}{\pvec a}{\pvec y} \cwedge \uInter{B}{\pvec b}{\pvec w}
    \using (\cwedge\textup{R})
\]
\justifies
\uInter{\Gamma}{\pvec u}{\pvec \gamma}, \uInter{\Delta}{\pvec v}{\pvec \delta}
    \proves \uInter{A \cwedge B}{\pvec a, \pvec b}{\pvec y, \pvec w}
\using (\hbox{Definition \ref{inter}})
\end{prooftree}
\quad \quad
\begin{prooftree}
\[
    \uInter{\Gamma}{\pvec u}{\pvec \gamma}, \uInter{A}{\pvec x}{\pvec a}, \uInter{B}{\pvec v}{\pvec b}
        \proves \uInter{C}{\pvec c}{\pvec w}
    \justifies
    \uInter{\Gamma}{\pvec u}{\pvec \gamma}, \uInter{A}{\pvec x}{\pvec a} \cwedge \uInter{B}{\pvec v}{\pvec b}
        \proves \uInter{C}{\pvec c}{\pvec w}
    \using (\cwedge\textup{L})
\]
\justifies
\uInter{\Gamma}{\pvec u}{\pvec \gamma}, \uInter{A \cwedge B}{\pvec x, \pvec v}{\pvec a, \pvec b}
    \proves \uInter{C}{\pvec c}{\pvec w}
\using (\hbox{Definition \ref{inter}})
\end{prooftree}
\]
\emph{$\lto\!\textup{L}$ introduction}.
\[
\begin{prooftree}
\[
    \[
        \uInter{\Gamma}{\pvec u}{\pvec \gamma[\pvec y]} \proves \uInter{A}{\pvec a}{\pvec y}
        \justifies
        \uInter{\Gamma}{\pvec u}{\pvec \gamma[\pvec f \pvec a (\pvec b[\pvec g \pvec a])]} \proves
               \uInter{A}{\pvec a}{\pvec f \pvec a (\pvec b[\pvec g \pvec a])}
   \using {[\frac{\pvec f \pvec a (\pvec b[\pvec g \pvec a])}{\pvec y}]}
    \]
    \quad
    \[
        \uInter{\Delta}{\pvec w}{\pvec \delta[\pvec v]}, \uInter{B}{\pvec v}{\pvec b[\pvec v]} \proves \uInter{C}{\pvec c[\pvec v]}{\pvec z}
        \justifies
        \uInter{\Delta}{\pvec w}{\pvec \delta[\pvec g \pvec a]}, \uInter{B}{\pvec g \pvec a}{\pvec b[\pvec g \pvec a]}
            \proves \uInter{C}{\pvec c[\pvec g \pvec a]}{\pvec z}
   \using {[\frac{\pvec g \pvec a}{\pvec v}]}
    \]
    \justifies
    \uInter{\Gamma}{\pvec u}{\pvec \gamma[\pvec f \pvec a (\pvec b[\pvec g \pvec a])]},
       \uInter{\Delta}{\pvec w}{\pvec \delta[\pvec g \pvec a]},
        \uInter{A}{\pvec a}{\pvec f \pvec a (\pvec b[\pvec g \pvec a])} \lto \uInter{B}{\pvec g \pvec a}{\pvec b[\pvec g \pvec a]}
            \proves \uInter{C}{\pvec c[\pvec g \pvec a]}{\pvec z}
    \using (\lto\textup{L})
\]
\justifies
\uInter{\Gamma}{\pvec u}{\pvec \gamma[\pvec f \pvec a (\pvec b[\pvec g \pvec a])]}, \uInter{\Delta}{\pvec w}{\pvec \delta[\pvec g \pvec a]},
    \uInter{A \lto B}{\pvec f, \pvec g}{\pvec a, \pvec b[\pvec g \pvec a]}
        \proves \uInter{C}{\pvec c[\pvec g \pvec a]}{\pvec z}
\using (\hbox{Definition \ref{inter}})
\end{prooftree}
\]
\emph{Universal quantifier}.
\[
\begin{prooftree}
\[
    \uInter{\Gamma}{\pvec u}{\pvec \gamma[z]}\proves \uInter{A(z)}{\pvec a[z]}{\pvec y}
    \justifies
    \uInter{\Gamma}{\pvec u}{\pvec \gamma[z]}\proves \uInter{A(z)}{(\lambda z.\pvec a[z])z}{\pvec y}
\]
\justifies
\uInter{\Gamma}{\pvec u}{\pvec \gamma[z]}\proves \uInter{\forall z A(z)}{\lambda z.\pvec a[z]}{\pvec y, z}
\using (\hbox{Definition \ref{inter}})
\end{prooftree}
\quad\quad\quad
\begin{prooftree}
\[
    \uInter{\Gamma}{\pvec u}{\pvec \gamma[\pvec x]}, \uInter{A(t)}{\pvec x}{\pvec
    a[\pvec x]}\proves \uInter{B}{\pvec b[\pvec x]}{\pvec w}
    \justifies
   \uInter{\Gamma}{\pvec u}{\pvec \gamma[\pvec f t]}, \uInter{A(t)}{\pvec f t}{\pvec
    a[\pvec f t]}\proves \uInter{B}{\pvec b[\pvec f t]}{\pvec w}
    \using [\frac{\pvec f t}{\pvec x}]
\]
\justifies
\uInter{\Gamma}{\pvec u}{\pvec \gamma[\pvec f t]}, \uInter{\forall z A(z)}{\pvec f}{\pvec
    a[\pvec f t], t}\proves \uInter{B}{\pvec b[\pvec f t]}{\pvec w}
\using (\hbox{Definition \ref{inter}})
\end{prooftree}
\]
\emph{Existential quantifier}.
\[
\begin{prooftree}
    \uInter{\Gamma}{\pvec u}{\pvec \gamma}\proves \uInter{A(t)}{\pvec a}{\pvec y}
    \justifies
    \uInter{\Gamma}{\pvec u}{\pvec \gamma}\proves \uInter{\exists z A(z)}{\pvec a,t}{\pvec y}
\using (\hbox{Definition \ref{inter}})
\end{prooftree}
\quad\quad\quad
\begin{prooftree}
    \uInter{\Gamma}{\pvec u}{\pvec \gamma[z]}, \uInter{A(z)}{\pvec x}{\pvec a[z]}\proves \uInter{B}{\pvec b[z]}{\pvec y}
    \justifies
   \uInter{\Gamma}{\pvec u}{\pvec \gamma[z]}, \uInter{\exists z A(z)}{\pvec x, z}{\pvec a[z]}\proves \uInter{B}{\pvec b[z]}{\pvec y}
\using (\hbox{Definition \ref{inter}})
\end{prooftree}
\]
\emph{$\&$\textup{R} introduction}.
\[
\begin{prooftree}
   \[
   \[
           \[
           \[
                \[
                	\uInter{\Gamma}{\pvec u}{\pvec \gamma_0 }\proves \uInter{A}{\pvec a}{\pvec y}
                	\quad
		\uInter{A}{\pvec a}{\pvec y} \proves \pcond{\true}{\uInter{A}{\pvec a}{\pvec y}}{\uInter{B}{\pvec b}{\pvec w}}
		\justifies
                   \uInter{\Gamma}{\pvec u}{\pvec \gamma_0} \proves 
                   \pcond{\true}{\uInter{A}{\pvec a}{\pvec y}}{\uInter{B}{\pvec b}{\pvec w}}
	       \]
                \justifies
                \proves \uInter{\Gamma}{\pvec u}{\pvec \gamma_0} \lto \pcond{\true}{\uInter{A}{\pvec a}{\pvec y}}{\uInter{B}{\pvec b}{\pvec w}}
                \]
                \justifies
                \proves \uInter{\Gamma}{\pvec u}{\true(\pvec \gamma_0,\pvec \gamma_1)}\lto
                   \pcond{\true}{\uInter{A}{\pvec a}{\pvec y}}{\uInter{B}{\pvec b}{\pvec w}}
          \using(\textup{Ax. 7})
            \]
           \quad \quad
           (+)
      \justifies
      \proves  \uInter{\Gamma}{\pvec u}{z(\pvec \gamma_0,\pvec \gamma_1)}\lto
             \pcond{z}{\uInter{A}{\pvec a}{\pvec y}}{\uInter{B}{\pvec b}{\pvec w}} \using(\&\textup{R})
       \using (\hbox{Lemma \ref{useful}(i)})
  \]
  \justifies
  \proves  \uInter{\Gamma}{\pvec u}{z(\pvec \gamma_0,\pvec
             \gamma_1)}\lto \uInter{A \awedge B}{\pvec a,\pvec b}{\pvec y,\pvec w,z}
   \using (\hbox{Definition \ref{inter}})
   \]
   \justifies
   \uInter{\Gamma}{\pvec u}{z(\pvec \gamma_0,\pvec
             \gamma_1)}\proves \uInter{A \awedge B}{\pvec a,\pvec b}{\pvec y,\pvec w,z}
\end{prooftree}
\]
where (+) is the dual case. \\[2mm]
\emph{$\&$\textup{L} introduction and $\oplus\textup{R}$ introduction}.
\[
\begin{prooftree}
\[
    \uInter{\Gamma}{\pvec u}{\pvec \gamma}, \uInter{A}{\pvec x}{\pvec a} \proves \uInter{B}{\pvec b}{\pvec w}
    \justifies
    \uInter{\Gamma}{\pvec u}{\pvec \gamma}, \pcond{\true}{\uInter{A}{\pvec x}{\pvec a}}{\uInter{C}{\pvec v}{\pvec c}}
       \proves \uInter{B}{\pvec b}{\pvec w}
    \using (\hbox{Lemma \ref{useful}(iv)})
\]
\justifies \uInter{\Gamma}{\pvec u}{\pvec \gamma}, \uInter{A \awedge C}{\pvec x,\pvec v}{\pvec a,\pvec c,\true}
      \proves \uInter{B}{\pvec b}{\pvec w}
\using (\hbox{Definition \ref{inter}})
\end{prooftree}
\quad\quad\quad
\begin{prooftree}
   \[
            \uInter{\Gamma}{\pvec u}{\pvec \gamma}\proves \uInter{A}{\pvec a}{\pvec y}
            \quad
             \uInter{A}{\pvec a}{\pvec y}\proves  \pcond{\true}{\uInter{A}{\pvec a}{\pvec y}}{\uInter{B}{\pvec b}{\pvec w}}
            \justifies \uInter{\Gamma}{\pvec u}{\pvec \gamma}\proves \pcond{\true}{\uInter{A}{\pvec a}{\pvec y}}{\uInter{B}{\pvec b}{\pvec w}} \using (cut)
            \]
  \justifies
   \uInter{\Gamma}{\pvec u}{\pvec \gamma}\proves \uInter{A\oplus B}{\pvec a,\pvec b,\true}{\pvec y,\pvec w}
\using (\hbox{Definition \ref{inter}})
\end{prooftree}
\]
The other $\&\textup{L}$ and $\oplus\textup{R}$ are similar. \\[2mm]
\emph{$\oplus\textup{L}$ introduction}.
\[
\begin{prooftree}
   \[
	(+)
           \quad
           \[
                   \uInter{\Gamma}{\pvec u}{\pvec \gamma_1}, \uInter{B}{\pvec v}{\pvec b}
                       \proves \uInter{C}{\pvec c_2}{\pvec w} \using[\frac{\pvec u}{\pvec r}]
                  \justifies
                \uInter{\Gamma}{\pvec u}{\false(\pvec \gamma_0,\pvec
                  \gamma_1)},\pcond{\false}{\uInter{A}{\pvec x}{\pvec
                    a}}{\uInter{B}{\pvec v}{\pvec b}}\proves
                \uInter{C}{\false(\pvec c_1,\pvec c_2)}{\pvec w}\using
                (\hbox{Ax.~ 7\, /\, Lemma \ref{useful} (iv)})
            \]
         \justifies
  \uInter{\Gamma}{\pvec u}{z(\pvec \gamma_0,\pvec \gamma_1)},
  \pcond{z}{\uInter{A}{\pvec x}{\pvec a}}{\uInter{B}{\pvec v}{\pvec
      b}}\proves  \uInter{C}{z(\pvec c_1,\pvec c_2)}{\pvec
    w}\using(\hbox{Lemma \ref{useful}(i)})
\]
\justifies
\uInter{\Gamma}{\pvec u}{z(\pvec \gamma_0,\pvec \gamma_1)}, \uInter{A\oplus B}{\pvec x,\pvec v,z}{\pvec a,\pvec b}  \proves \uInter{C}{z(\pvec c_1,\pvec c_2)}{\pvec w}
\using (\hbox{Definition \ref{inter}})
\end{prooftree}
\]
where (+) is the dual case. The other rules are treated similarly.\qed

\subsection{Characterisation}
\label{sec:characterisation}

As mentioned in the introduction, one of the main advantages of working in the context of intuitionistic linear logic is that we no longer need branching quantifiers. The asymmetry introduced in $\ILLomega$ turns the symmetric games of classical linear logic into games where Eloise always plays first, so formulas $A$ are interpreted as $\exists \pvec x \forall \pvec y \uInter{A}{\pvec x}{\pvec y}$.

\begin{prop} The following principles, denoted by $\lAC$, $\lMP$, $\lIP$ and $\EP$ (acronyms for linear versions of Axiom of Choice, Markov Principle, Independence of Premises and Extra Principle) characterise the basic interpretation presented above
\eqleft{
\begin{array}{lcl}
\lAC & \; \colon \; & \forall \pvec x \exists \pvec y A_{\forall}(\pvec y) \lto \exists \pvec f \forall \pvec x A_{\forall}(\pvec f \pvec x) \\[2mm]
\lMP & \colon & (\forall \pvec x \Aqf \lto \Bqf) \lto \exists \pvec x (\Aqf \lto \Bqf) \\[2mm]
\lIP & \colon & (A_{\forall} \lto \exists \pvec y B_{\forall}) \lto \exists \pvec y (A_{\forall} \lto B_{\forall}) \\[2mm]
\EP & \colon & \forall \pvec x, \pvec v (\Aqf \cwedge \Bqf) \lto (\forall \pvec x \Aqf \cwedge \forall \pvec v \Bqf) \\[2mm]
\end{array}
}
where $\Aqf$, $\Bqf$ and $A_{\forall}$, $B_{\forall}$ are quantifier-free formulas and purely universal formulas of $\ILLomegab$ respectively. It is also assumed that $\pvec x$ does not occur in $\Bqf$, $\pvec y$ does not occur in $A_{\forall}$ in the principle $\lIP$ and $\pvec v$ does not occur in $\Aqf$ in the principle $\EP$. Formally,
\eqleft{\ILLomegab + \lAC + \lMP + \lIP + \EP \proves A \lequiv \exists \pvec x \forall \pvec y \uInter{A}{\pvec x}{\pvec y}.} Moreover, assuming that $\pcond{}{}{}$ is a primitive symbol, interpreted as in \cite{Oliva(2007)}, the characterisation result still holds when bang does not occur in $\lAC$, $\lMP$, $\lIP$ and $\EP$ and these principles are interpretable, i.e. denoting by $P$ any instance of these principles, there are terms $\pvec a$ such that $\ILLomegab\proves \forall \pvec y\uInter{P}{\pvec a}{\pvec y}$.
\end{prop}
\proof The linear equivalence can be proved by induction on the logical structure of $A$. Let us consider a few cases: \\[2mm]
\emph{Tensor}.
\eqleft{
\begin{array}{lcl}
A \cwedge B & \stackrel{(\textup{IH})}{\lequiv} & \exists \pvec x \forall \pvec y \uInter{A}{\pvec x}{\pvec y} \cwedge \exists \pvec v \forall \pvec w \uInter{B}{\pvec v}{\pvec w} \\[2mm]
  & \stackrel{(\EP)}{\lequiv} & \exists \pvec x, \pvec v \forall \pvec y, \pvec w (\uInter{A}{\pvec x}{\pvec y} \cwedge \uInter{B}{\pvec v}{\pvec w}) \\[2mm]
  & \equiv & \exists \pvec x, \pvec v \forall \pvec y, \pvec w \uInter{A \cwedge B}{\pvec x, \pvec v}{\pvec y, \pvec w}.
\end{array}
}
\emph{With}.
\eqleft{
\begin{array}{lcl}
A \awedge B & \stackrel{(\textup{IH})}{\lequiv} & \exists \pvec x \forall \pvec y \uInter{A}{\pvec x}{\pvec y} \awedge \exists \pvec v \forall \pvec w \uInter{B}{\pvec v}{\pvec w} \\[2mm]
  & \lequiv & \forall z (\pcond{z}{\exists \pvec x \forall \pvec y \uInter{A}{\pvec x}{\pvec y}}{ \exists \pvec v \forall \pvec w \uInter{B}{\pvec v}{\pvec w}}) \\[2mm]
  & \lequiv & \forall z \exists \pvec x, \pvec v(\pcond{z}{ \forall \pvec y \uInter{A}{\pvec x}{\pvec y}}{ \forall \pvec w \uInter{B}{\pvec v}{\pvec w}}) \\[2mm]
  & \lequiv & \forall z \exists \pvec x, \pvec v\forall \pvec y, \pvec w(\pcond{z}{\uInter{A}{\pvec x}{\pvec y}}{\uInter{B}{\pvec v}{\pvec w}}) \\[2mm]
  & \stackrel{(\lAC)}{\lequiv} & \exists \pvec f, \pvec g\forall z, \pvec y, \pvec w(\pcond{z}{\uInter{A}{\pvec f z}{\pvec y}}{\uInter{B}{\pvec g z}{\pvec w}}) \\[2mm]
   & \lequiv & \exists \pvec x, \pvec v\forall z, \pvec y, \pvec w(\pcond{z}{\uInter{A}{\pvec x}{\pvec y}}{\uInter{B}{\pvec v}{\pvec w}}) \\[2mm]
  & \equiv & \exists \pvec x, \pvec v \forall \pvec y, \pvec w, z \uInter{A \awedge B}{\pvec x, \pvec v}{\pvec y, \pvec w,z}.
\end{array}
}
\emph{Linear implication}.
\eqleft{
\begin{array}{lcl}
A \lto B & \stackrel{(\textup{IH})}{\lequiv} & \exists \pvec x \forall \pvec y \uInter{A}{\pvec x}{\pvec y} \lto \exists \pvec v \forall \pvec w \uInter{B}{\pvec v}{\pvec w}
  \stackrel{(\lIP, \lMP)}{\lequiv} \forall \pvec x \exists \pvec v \forall \pvec w \exists \pvec y (\uInter{A}{\pvec x}{\pvec y} \lto \uInter{B}{\pvec v}{\pvec w}) \\[2mm]
  & \stackrel{(\lAC)}{\lequiv} & \exists \pvec f, \pvec g \forall \pvec x, \pvec w (\uInter{A}{\pvec x}{\pvec f \pvec x \pvec w} \lto \uInter{B}{\pvec g \pvec x}{\pvec w})
  \equiv \exists \pvec f, \pvec g \forall \pvec x, \pvec w \uInter{A \lto B}{\pvec f, \pvec g}{\pvec x, \pvec w}.
\end{array}
}
\emph{Universal quantifier}.
\eqleft{
\forall z A \stackrel{(\textup{IH})}{\lequiv} \forall z \exists \pvec x \forall \pvec y \uInter{A}{\pvec x}{\pvec y}
  \stackrel{(\lAC)}{\lequiv} \exists \pvec f \forall \pvec y, z \uInter{A}{\pvec f z}{\pvec y}
  \equiv \exists \pvec f \forall \pvec y, z \uInter{\forall z A}{\pvec f}{\pvec y, z}.
}
The other cases are treated similarly. In fact, for the remaining
cases (once the induction hypothesis is assumed) the equivalence can
be proved in $\ILLomegab$ alone. 

 With the assumptions presented, the interpretability of the principles is easily checked since quantifier-free formulas are interpretable by themselves, i.e. they do not ask for realisers. We illustrate with the principle $\lAC$ where the premise is interpreted as
\[ \uInter{\forall \pvec x \exists \pvec y \forall \pvec z A_{qf}}{\pvec f}{\pvec z, \pvec x}\equiv \uInter{\exists \pvec y \forall \pvec z A_{qf}(\pvec x, \pvec y, \pvec z)}{\pvec f \pvec x}{\pvec z}\equiv \uInter{\forall \pvec z A_{qf}(\pvec x, \pvec f\pvec x, \pvec z)}{}{\pvec z}\equiv A_{qf}(\pvec x, \pvec f\pvec x, \pvec z) \]
whereas the conclusion is interpreted as
\[ \uInter{\exists \pvec f \forall \pvec x \forall \pvec z A_{qf}(\pvec x, \pvec f \pvec x, \pvec z)}{\pvec f}{\pvec z,\pvec x}\equiv \uInter{\forall \pvec x \forall \pvec z A_{qf}(\pvec x, \pvec f \pvec x, \pvec z)}{}{\pvec z,\pvec x}\equiv\uInter{\forall \pvec z A_{qf}(\pvec x, \pvec f \pvec x, \pvec z)}{}{\pvec z}\equiv A_{qf}(\pvec x, \pvec f \pvec x, \pvec z). \]
Since the realisers of the premise are the same as those of the conclusion, the identity and projection functions can be taken as realisers of the implication. \qed

\begin{rem} Note that if we are embedding $\ILomega$ via the standard embedding $\lTrans{(\cdot)}$ then the connective $ \cwedge $ is not needed, and hence the extra principle $\EP$ is not needed either.
\end{rem}

\section{Some Interpretations of  \texorpdfstring{$\ILLomega$}{ILLomega}}
\label{sec:modal}

In this section we consider a few choices of how the basic interpretation given in Definition \ref{inter} can be extended to full intuitionistic linear logic, i.e. we present three possible interpretations of $\bang A$. All choices considered will have the form
\begin{equation} \label{general}
\uInter{\bang A}{\pvec x}{\pvec y} \pdefin \,\bang \ubq{\pvec y'}{\pvec y}{\uInter{A}{\pvec x}{\pvec y'}}
\end{equation}
where $\ubq{\pvec y}{\pvec a}{A}$ is a meta-level formula construction which we will assume to satisfy the following: For some terms\footnote{Note that these terms are allowed to be specific to the formula $A$, in particular, the free variables of $\singleton{(\cdot)}, \join{(\cdot)}{(\cdot)}$ and $\comp{(\cdot)}{(\cdot)}$ are assumed to be contained in the free-variables of $\forall \pvec y A[\pvec y]$ (i.e. all free-variables of $A$ except $\pvec y$).}  $\singleton{(\cdot)}, \join{(\cdot)}{(\cdot)}$ and $\comp{(\cdot)}{(\cdot)}$ the conditions below are provable in $\ILLomegab$
\begin{itemize}
\item[\textup{(A1)}] $\bang \ubq{\pvec y}{\singleton{(\pvec z)}}{A[\pvec y]} \lto A[\pvec z]$
\item[\textup{(A2)}] $\bang \ubq{\pvec y}{(\join{\pvec y_1}{\pvec y_2})}{A[\pvec y]} \lto \,\bang (\ubq{\pvec y}{\pvec y_1}{A[\pvec y]}) \,\cwedge\; \bang (\ubq{\pvec y}{\pvec y_2}{A[\pvec y]})$
\item[\textup{(A3)}] $\bang \ubq{\pvec y}{(\comp{\pvec f}{\pvec z})}{A[\pvec y]} \lto \, \bang \ubq{\pvec x}{\pvec z}{\bang \ubq{\pvec y}{\pvec f \pvec x}{A[\pvec y]}}$.
\end{itemize}

The three instances of such meta-level formula construction $\ubq{\pvec y}{\pvec a}{A}$ we will consider are $\forall \pvec y A$, $\forall \pvec y \!\in\! \pvec a \, A$ (where $\pvec y \!\in\! \pvec a$ will be defined later), and $A[\pvec a / \pvec y]$.

\begin{prop}\label{generalbang} Under the assumptions (A1 -- A3) on the formula construction $\ubq{\pvec y}{\pvec a}{A}$, the generic interpretation of $\bang A$ as above leads to a sound functional interpretation of $\ILLomega$.
\end{prop}
\proof By Theorem \ref{soundness} we just have to analyse the rules of contraction, weakening, $\bang\textup{R}$, and $\bang\textup{L}$.\medskip

\noindent\emph{Contraction}. Assume by induction hypothesis that we already have terms witnessing the premise of the rule, i.e. $\uInter{\Gamma}{\pvec u}{\pvec \gamma}, \uInter{\bang A}{\pvec x_0}{\pvec a_0}, \uInter{\bang A}{\pvec x_1}{\pvec a_1}\proves \uInter{B}{\pvec b}{\pvec w}$. We must from these construct witnesses for the conclusion $\Gamma, \bang A \proves B$. That can be done as follows: 
\[
\begin{prooftree}
\[
\[
  \[
      \[
      \uInter{\Gamma}{\pvec u}{\pvec \gamma}, \uInter{\bang A}{\pvec x_0}{\pvec a_0}, \uInter{\bang A}{\pvec x_1}{\pvec a_1}\proves \uInter{B}{\pvec b}{\pvec w}
      \justifies
       \uInter{\Gamma}{\pvec u}{\pvec \gamma}, \uInter{\bang A}{\pvec x}{\pvec a_0}, \uInter{\bang A}{\pvec x}{\pvec a_1}\proves \uInter{B}{\pvec b}{\pvec w}
       \using[\frac{\pvec x}{\pvec x_0}, \frac{\pvec x}{\pvec x_1}]
      \]
      \justifies
      \uInter{\Gamma}{\pvec u}{\pvec \gamma}, \bang \ubq{\pvec y'}{\pvec a_0}{\uInter{A}{\pvec x}{\pvec y'}}, \bang \ubq{\pvec y'}{\pvec a_1}{\uInter{A}{\pvec x}{\pvec y'}} \proves \uInter{B}{\pvec b}{\pvec w}
       \using {(\ref{general})}
  \]
  \justifies
  \uInter{\Gamma}{\pvec u}{\pvec \gamma}, \bang \ubq{\pvec y'}{\pvec a_0}{\uInter{A}{\pvec x}{\pvec y'}} \cwedge~ \bang \ubq{\pvec y'}{\pvec a_1}{\uInter{A}{\pvec x}{\pvec y'}} \proves \uInter{B}{\pvec b}{\pvec w}
    \using {(\otimes\textup{L})}
\]
\justifies
\uInter{\Gamma}{\pvec u}{\pvec \gamma}, \bang \ubq{\pvec y'}{\join{\pvec a_0}{\pvec a_1}}{\uInter{A}{\pvec x}{\pvec y'}} \proves \uInter{B}{\pvec b}{\pvec
       w}\using(\textup{A}2)
\]
\justifies \uInter{\Gamma}{\pvec u}{\pvec \gamma}, \uInter{\bang A}{\pvec x}{\join{\pvec a_0}{\pvec a_1}}\proves \uInter{B}{\pvec
b}{\pvec w}
\using {(\ref{general})}
\end{prooftree}
\]
\emph{Weakening}.
\[
\begin{prooftree}
      \[
      \uInter{\Gamma}{\pvec u}{\pvec \gamma}\proves \uInter{B}{\pvec b}{\pvec w}
      \justifies
       \uInter{\Gamma}{\pvec u}{\pvec \gamma}, \bang \ubq{\pvec y'}{\pvec a}{\uInter{A}{\pvec x}{\pvec y'}} \proves \uInter{B}{\pvec b}{\pvec w}
       \using(\textup{wkn})
      \]
      \justifies
      \uInter{\Gamma}{\pvec u}{\pvec \gamma}, \uInter{\bang A}{\pvec x}{\pvec a}\proves \uInter{B}{\pvec b}{\pvec w}
       \using {(\ref{general})}
\end{prooftree}
\]
where $\pvec a$ are arbitrary closed terms of the appropriate types. Note that every type is inhabited by a closed term.\medskip

\noindent\emph{$\bang\textup{R}$}.
\[
\begin{prooftree}
\[
  \[
      \[
      \uInter{\bang \Gamma}{\pvec u}{\pvec \gamma[\pvec y']}\proves \uInter{A}{\pvec a}{\pvec y'}
      \justifies
       \bang \ubq{\pvec w'}{\pvec \gamma [\pvec y']}{\uInter{\Gamma}{\pvec u}{\pvec w'}} \proves \uInter{A}{\pvec a}{\pvec y'}
       \using {(\ref{general})}
      \]
      \justifies
      \bang \ubq{\pvec y'}{\pvec y}{\bang \ubq{\pvec w'}{(\lambda \pvec y'.\pvec \gamma [\pvec y'])\pvec y'}{\uInter{\Gamma}{\pvec u}{\pvec w'}}}
       \proves \,\bang \ubq{\pvec y'}{\pvec y}{\uInter{A}{\pvec a}{\pvec y'}}
  \]
  \justifies
 \bang \ubq{\pvec w'}{\comp{(\lambda \pvec y' .\pvec \gamma[\pvec y'])}{\pvec y}}{\uInter{\Gamma}{\pvec u}{\pvec w'}} \proves \, \bang \ubq{\pvec y'}{\pvec y}{\uInter{A}{\pvec a}{\pvec y'}}
 \using(\textup{A}3)
\]
\justifies \uInter{\bang \Gamma}{\pvec u}{\comp{(\lambda \pvec y' .\pvec \gamma[\pvec y'])}{\pvec y}}\proves \uInter{\bang A}{\pvec a}{\pvec y}
\using {(\ref{general})}
\end{prooftree}
\]
\emph{$\bang\textup{L}$}.
\[
\begin{prooftree}
\[
           \uInter{\Gamma}{\pvec u}{\pvec \gamma}, \uInter{A}{\pvec x}{\pvec a}\proves \uInter{B}{\pvec b}{\pvec w}
   \justifies
   \uInter{\Gamma}{\pvec u}{\pvec \gamma}, \bang \ubq{\pvec y}{\singleton{(\pvec a)}}{\uInter{A}{\pvec x}{\pvec y}}
       \proves \uInter{B}{\pvec b}{\pvec w}
   \using(\textup{A}1)
\]
\justifies
\uInter{\Gamma}{\pvec u}{\pvec \gamma}, \uInter{\bang A}{\pvec x}{\singleton{(\pvec a)}}\proves \uInter{B}{\pvec b}{\pvec w}
\using {(\ref{general})}
\end{prooftree}
\]
That concludes the proof. \qed

\begin{rem} Assume that the types of $\pvec y^{\rho}$ and ${\pvec a}^{T \rho}$ in $\ubq{\pvec y}{\pvec a}{\uInter{A}{\pvec x}{\pvec y}}$ are as shown, for a fixed $A$. Then, our three families of terms have types
\eqleft{
\begin{array}{lcl}
   \singleton{} & \colon & \rho \to T \rho \\[2mm]
   \join{}{} & \colon & T \rho \times T \rho \to T \rho \\[2mm]
   \comp{}{} & \colon & (\tau \to T \rho) \times T \tau \to T \rho.
\end{array}
}
In category theory, one could think of $(T, \singleton{}, \comp{}{})$ as forming a Kleisli triple ($\sim$ monad), with $\join{}{}$ being a commutative monoid on $T \rho$. This in turn extends to a comonad on formulas as
\eqleft{T(A[\pvec y]) \pdefin \, \bang \ubq{\pvec y}{\pvec a}{A},}
\noindent where the formula $A$ with free-variables $\pvec y$ is transformed in the new formula $\bang \ubq{\pvec y}{\pvec a}{A}$ with free-variables $\pvec a$.
See e.g. the work of Valeria de Paiva \cite{dePaiva(1989B)} and Martin Hyland (\cite{Hyland(02)}, section 3.1) on categorical logic for more information about the connection between functional interpretations and comonads. More on the relation between ours and de Paiva's work can be found in Section \ref{categorical}.
\end{rem}

Next, we present three sound interpretations of $\bang A$ by providing three instances of $\ubq{\pvec y}{\pvec a}{A}$ which satisfy conditions (A1), (A2), and (A3). It is important to observe that the meta-level formula construction $\ubq{\pvec y}{\pvec a}{A}$ is part of the \emph{verifying system}. Therefore, when discussing particular instances of $\ubq{\pvec y}{\pvec a}{A}$ it is not relevant for the interpretation how the terms needed are axiomatised. Only axioms in the \emph{interpreted system} needed to be interpreted.

\begin{prop} \label{banginter} We have the following:
\begin{enumerate}[\em(a)]
\item $\uInter{\bang A}{\pvec x}{}  \pdefin \, \bang \forall \pvec y \uInter{A}{\pvec x}{\pvec y} $ is a sound interpretation of $\bang A$.

\item Assume that the language of the verifying system $\ILLomegab$
  has a new finite type $\sigma^*$ for each finite type $\sigma$. An
  element of type $\sigma^*$ is a finite set of elements of type
  $\sigma$. The extended language has a relation symbol $\in$ infixing
  between a term of type $\sigma$ and a term of type $\sigma^*$ with
  axioms to ensure that $\bang (x\in y)$ if and only if $x$ is an
  element in the set $y$. Let then the formula $\forall \pvec x
  \!\in\! \pvec t \, A$ abbreviate $\forall \pvec x (\bang (\pvec x
  \in \pvec t) \lto A)$.  Assume also the existence of three more
  constants $\singleton{}:\sigma\to\sigma^*$,
  $\join{}{}:\sigma^*\to\sigma^*\to\sigma^*$ and
  $\comp{}{}:\sigma^*\to(\sigma\to\rho^*)\to\rho^*$ that should be
  seen as terms such that $\singleton{(t)}$ is the singleton set with
  $t^{\sigma}$ as the only element (in particular $\bang (t\in
  \singleton{(t)})$), $\join{t}{q}$ is the union of two finite sets
  $t$ and $q$, and $\comp{f}{q}$ is the set that results from the
  union of all sets $fx$ with $x\in q$. Then $\uInter{\bang A}{\pvec
    x}{\pvec a} \pdefin \,\bang \forall \pvec y \!\in\! \pvec a \,
  \uInter{A}{\pvec x}{\pvec y}$ is a sound interpretation of $\bang
  A$.
\item  Assume the verifying system $\ILLomegab$ has an extra axiom schema $\vdash \, \bang A \oplus (\bang A\lto 0)$, asserting the decidability of quantifier free-formulas $A$. Assume also that definition by cases is definable over quantifier-free formulas $A$ in the term language of $\ILLomegab$, i.e.
\eqleft{\join{\pvec t}{\pvec s} :=
\left\{
\begin{array}{ll}
 \pvec t \quad & {\sf if} \; \bang A \lto 0 \\[2mm]
 \pvec s & {\sf if} \; \bang A,
\end{array}
\right.
}
with the rules
\[
\begin{array}{llr}
\begin{prooftree} \Gamma \proves B[\join{\pvec t}{\pvec s}] \justifies \Gamma, \bang A \proves B[\pvec s]
\end{prooftree} & \quad \quad \quad &
\begin{prooftree} \Gamma \proves B[\join{\pvec t}{\pvec s} ]  \justifies \Gamma, \bang A \lto 0 \proves B[\pvec t]
\end{prooftree}
\end{array}
\]
Then, $\uInter{\bang A}{\pvec x}{\pvec y}  \pdefin \, \bang \uInter{A}{\pvec x}{\pvec y}$ is a sound interpretation of $\bang A$.
\end{enumerate}
\end{prop}
\proof \hfill
\begin{enumerate}[(a)]
\item This interpretation of $\bang A$ corresponds to the choice
$\ubq{\pvec y}{\pvec t}{A} \pdefin \forall \pvec y A$. It is easy to check that conditions $(A1)$, $(A2)$ and $(A3)$ become
\eqleft{
\begin{array}{l}
\bang \forall \pvec y  A[\pvec y] \lto A[\pvec z] \\[2mm]
\bang \forall \pvec y A[\pvec y] \lto \,\bang \forall \pvec y A[\pvec y] \,\cwedge\; \bang \forall \pvec y A[\pvec y] \\[2mm]
\bang \forall \pvec y A[\pvec y] \lto \,\bang \forall \pvec x  \bang \forall \pvec y A[\pvec y]
\end{array}
}
respectively, which are trivially derivable in $\ILLomegab$.

\item The interpretation $\uInter{\bang A}{\pvec x}{\pvec a}  \pdefin \,
\bang \forall \pvec y \!\in\! \pvec a \, \uInter{A}{\pvec x}{\pvec
y}$ corresponds to the choice $\ubq{\pvec y}{\pvec t}{A} \pdefin \forall \pvec y \!\in\! \pvec t \, A$, i.e. $\forall \pvec y (\bang (\pvec y\in \pvec t)
\lto A[\pvec y])$.
In this context, the conditions $(A1)$, $(A2)$ and $(A3)$ become
\eqleft{
\begin{array}{l}
\bang \forall \pvec y \!\in\! \singleton{(\pvec z)} \, A[\pvec y] \lto A[\pvec z] \\[2mm]
\bang \forall \pvec y \!\in\! \join{\pvec y_1}{\pvec y_2} \, A[\pvec y] \lto \,\bang \forall \pvec y \!\in\! \pvec y_1 \, A[\pvec y] \,\cwedge\; \bang \forall \pvec y \!\in\! \pvec y_2 \, A[\pvec y] \\[2mm]
\bang \forall \pvec y \! \in\! \comp{\pvec f}{\pvec z} \, A[\pvec y] \lto \, \bang \forall \pvec x \!\in\! \pvec z ~ \bang \forall \pvec y \!\in\! \pvec f \pvec x A[\pvec y],
\end{array}
}
which are derivable in the extension of $\ILLomegab$ outlined above.

\item This interpretation of $\bang A$ corresponds to the choice $\ubq{\pvec y}{\pvec t}{A[\pvec y]} \pdefin A[\pvec t/\pvec y]$. Given a formula $A[\pvec y]$ we define $\singleton{(\cdot)}$, as being the identity, $\comp{}{}$ is defined as $\comp{f}{x} \pdefin f x$ and $\join{\pvec y_1}{\pvec y_2}$ as
\eqleft{\join{\pvec y_1}{\pvec y_2} :=
\left\{
\begin{array}{ll}
 \pvec y_1 \quad & {\sf if} \; \bang A[\pvec y_1] \lto 0 \\[2mm]
 \pvec y_2 & {\sf if} \; \bang A[\pvec y_1].
\end{array}
\right.
}
Conditions $(A1)$, $(A2)$ and $(A3)$ then become
\eqleft{
\begin{array}{l}
\bang A[\singleton{(\pvec z)}] \lto A[\pvec z] \\[2mm]
\bang A[\join{\pvec y_1}{\pvec y_2}] \lto \,\bang A[\pvec y_1] \,\cwedge\; \bang A[\pvec y_2] \\[2mm]
\bang A[\comp{\pvec f}{{\pvec z}}] \lto \, \bang \bang A[\pvec f \pvec z]
\end{array}
}
respectively. From the definitions of $\singleton{(\cdot)}$ and $\comp{(\cdot)}{(\cdot)}$ conditions $(A1)$ and $(A3)$ are trivially derivable. In the derivation of
$(A2)$ we use
\begin{itemize}
   \item[] $\vdash ~\bang A \oplus (\bang A\lto 0)$
   \item[] $\bang A[\pvec y_1], \bang A[\join{\pvec y_1}{\pvec y_2}] \vdash \, \bang A[\pvec y_1] \,\cwedge \,\bang A[\pvec y_2]$, and
   \item[] $\bang A[\pvec y_1]\lto 0, \bang A[\join{\pvec y_1}{\pvec y_2}]\vdash 0$.
\end{itemize}
More precisely,
{\small\[
\begin{prooftree}
  (+)
  \quad
  \[
  \bang A[\pvec y_1], \bang A[\join{\pvec y_1}{\pvec y_2}]\proves ~\bang A[\pvec y_1]\otimes \bang A[\pvec y_2]
  \quad
  \[
  \bang A[\pvec y_1]\lto 0, \bang A[\join{\pvec y_1}{\pvec y_2}]\proves 0
  \justifies
  \bang A[\pvec y_1]\lto 0, \bang A[\join{\pvec y_1}{\pvec y_2}]\proves ~\bang A[\pvec y_1]\otimes \bang A[\pvec y_2] \using{(cut)}
  \]
  \justifies
  \bang A[\pvec y_1] \oplus (\bang A[\pvec y_1]\lto 0), \bang A[\join{\pvec y_1}{\pvec y_2}]\proves ~\bang A[\pvec y_1]\otimes \bang A[\pvec y_2]
  \]
  \justifies
  \bang A[\join{\pvec y_1}{\pvec y_2}]\proves ~\bang A[\pvec y_1]\otimes \bang A[\pvec y_2]
\end{prooftree}
\]}
where $(+)$ is an instance of the assumed axiom $\bang A[\pvec y_1] \oplus (\bang A[\pvec y_1]\lto 0)$. \qed
\end{enumerate}

\section{Relation to Standard Interpretations of \texorpdfstring{$\ILomega$}{ILomega}}
\label{sec:relation}

We argued in the introduction (see Proposition \ref{g-trans}) that for the purpose of analysing $\ILomega$ via linear logic it suffices to work with the system $\ILLomegar$. As it turns out, in $\ILLomegar$, we can simplify the interpretation of the connective $\awedge$, so that we no longer need the boolean variable $z$ in $\Diamond_{z}$ in that particular case.

\begin{prop} \label{main-prop} When interpreting the subsystem $\ILLomegar$, the interpretation of $A \awedge B$ presented in Definition \ref{inter} can be simplified so that the parametrised interpretation
\eqleft{
\begin{array}{lcl}
\uInter{A \lto B}{\pvec f, \pvec g}{\pvec x, \pvec w} & \pdefin & \uInter{A}{\pvec x}{\pvec f \pvec x \pvec w} \lto \uInter{B}{\pvec g \pvec x}{\pvec w} \\[2mm]
\uInter{A \cwedge B}{\pvec x, \pvec v}{\pvec y, \pvec w} & \pdefin & \uInter{A}{\pvec x}{\pvec y} \cwedge \uInter{B}{\pvec v}{\pvec w} \\[2mm]
\uInter{A \awedge B}{\pvec x, \pvec v}{\pvec y, \pvec w} & \pdefin & \uInter{A}{\pvec x}{\pvec y} \awedge \uInter{B}{\pvec v}{\pvec w} \\[2mm]
\uInter{A \oplus B}{\pvec x, \pvec v,z}{\pvec y, \pvec w} & \pdefin & \pcond{z}{\uInter{A}{\pvec x}{\pvec y}}{\uInter{B}{\pvec v}{\pvec w}} \\[2mm]
\uInter{\exists z A(z)}{\pvec x, z}{\pvec y} & \pdefin & \uInter{A(z)}{\pvec x}{\pvec y} \\[2mm]
\uInter{\forall z A(z)}{\pvec f}{\pvec y, z} & \pdefin & \uInter{A(z)}{\pvec f z}{\pvec y} \\[2mm]
\uInter{\bang A}{\pvec x}{\pvec y} & \pdefin & \bang \ubq{\pvec y'}{\pvec y}{\uInter{A}{\pvec x}{\pvec y'}}
\end{array}}
is sound for $\ILLomegar$, assuming \textup{(A1)}, \textup{(A2)}, and \textup{(A3)} are satisfied.
\end{prop}
\proof We just have to analyse the rules for $\awedge$ having in mind that, in the
case of the system under interpretation, the $\&\textup{R}$ introduction is restricted of the form $\bang \Gamma$. The simplified interpretation of $A \awedge B$ is shown sound as:
\[
\begin{prooftree}
\[
   \[
       \[
           \uInter{\bang \Gamma}{\pvec u}{\pvec \gamma_0}\proves \uInter{A}{\pvec a}{\pvec x}
           \justifies
           \bang \ubq{\pvec y'}{\pvec \gamma_0}{\uInter{\Gamma}{\pvec u}{\pvec y'}} \proves \uInter{A}{\pvec a}{\pvec x}
           \using {(\textup{P}\ref{main-prop})}
       \]
       \justifies
       \bang \ubq{\pvec y'}{(\join{\pvec \gamma_0}{\pvec \gamma_1})}{\uInter{\Gamma}{\pvec u}{\pvec y'}}
           \proves \uInter{A}{\pvec a}{\pvec x}
       \using {(\textup{A}2)}
   \]
   \quad
   \[
       \[
           \uInter{\bang\Gamma}{\pvec u}{\pvec \gamma_1}\proves \uInter{B}{\pvec b}{\pvec y}
           \justifies
           \bang \ubq{\pvec y'}{\pvec \gamma_1}{\uInter{\Gamma}{\pvec u}{\pvec y'}} \proves \uInter{B}{\pvec b}{\pvec y}                       \using {(\textup{P}\ref{main-prop})}
       \]
               \justifies
       \bang \ubq{\pvec y'}{(\join{\pvec \gamma_0}{\pvec \gamma_1})}{\uInter{\Gamma}{\pvec u}{\pvec y'}}
           \proves \uInter{B}{\pvec b}{\pvec y}
       \using {(\textup{A}2)}
   \]
   \justifies
   \bang \ubq{\pvec y'}{(\join{\pvec \gamma_0}{\pvec \gamma_1})}{\uInter{\Gamma}{\pvec u}{\pvec y'}}
       \proves \uInter{A}{\pvec a}{\pvec x} \awedge  \uInter{B}{\pvec b}{\pvec y} \using(\&\textup{R})
\]
\justifies
\uInter{\bang\Gamma}{\pvec u}{\join{\pvec \gamma_0}{\pvec \gamma_1}}\proves \uInter{A \awedge B}{\pvec a,\pvec b}{\pvec x,\pvec y}
\using {(\textup{P}\ref{main-prop})}
\end{prooftree}
\]
And for the left introduction:
\[
\begin{prooftree}
\[
   \uInter{\Gamma}{\pvec u}{\pvec \gamma},\uInter{A}{\pvec x}{\pvec a}\proves \uInter{C}{\pvec c}{\pvec w}
   \justifies
   \uInter{\Gamma}{\pvec u}{\pvec \gamma},  \uInter{A}{\pvec x}{\pvec a} \awedge  \uInter{B}{\pvec v}{\pvec b}
       \proves \uInter{C}{\pvec c}{\pvec w} \using (\&\textup{L})
\]
\justifies \uInter{\Gamma}{\pvec u}{\pvec \gamma}, \uInter{A \awedge B}{\pvec x,\pvec v}{\pvec a,\pvec b}
   \proves \uInter{C}{\pvec c}{\pvec w}
\using {(\textup{P}\ref{main-prop})}
\end{prooftree}
\]
The other $\&\textup{L}$ introduction is similar.\qed

Since in the remaining part of this section we work with translations of intuitionistic logic into linear logic, by $\uInter{A}{\pvec x}{\pvec y}$ we refer to the (simplified) parametrised interpretation described in Proposition \ref{main-prop}. Next we prove that the three different ways of interpreting $\bang A$ (cf. Proposition \ref{banginter}) give rise to interpretations of $\ILLomegar$ that correspond (via the translations of intuitionistic logic into intuitionistic linear logic) to Kreisel's modified realizability, the Diller-Nahm interpretation, and G\"{o}del's Dialectica interpretation, as:
\[
\begin{array}{l|cl}
\uInter{\bang A}{\pvec x}{\pvec a} & & \mbox{Interpretation of $\ILomega$} \\[1mm]
\hline \\[-2mm]
\bang \forall \pvec y \uInter{A}{\pvec x}{\pvec y} & & \textup{Kreisel modified realizability} \\[2mm]
\bang \forall \pvec y \!\in\! \pvec a \, \uInter{A}{\pvec x}{\pvec y} \;\; & & \textup{Diller-Nahm interpretation} \\[2mm]
\bang \uInter{A}{\pvec x}{\pvec a} & & \textup{G\"{o}del's Dialectica interpretation}.
\end{array}
\]
But first we introduce a simplified version of the translation $\lTrans{(\cdot)}$ from $\ILomega$ into $\ILLomegar$, which we will use in the treatment of the Diller-Nahm and the Dialectica interpretations (for modified realizability we use the translation $\bTrans{(\cdot)}$). This simplification of Girard's translation is necessary so as to obtain an exact match between intuitionistic and linear interpretations. The simplification, however, requires two additional principles which, as we will see, turn out to be interpretable.

\begin{prop} \label{simp-emb} Consider the following simplification of Girard's translation $\lTrans{(\cdot)}$, where the translation of $\vee$ and $\exists$ no longer needs the introduction of $\bang$  (cf. Proposition \ref{g-trans})
\eqleft{
\begin{array}{ll}
\pTrans{\Aat} & \pdefin  \Aat, \quad \text{ if } \Aat \not\equiv \bot \\[2mm]
\pTrans{\bot}          &\pdefin 0              \\[2mm]
\pTrans{(A \wedge B)}     &\pdefin \pTrans{A} \awedge \pTrans{B} \\[2mm]
\pTrans{(A \vee B)}     &\pdefin \pTrans{A} \oplus \pTrans{B} \\[2mm]
\pTrans{(A \to B)}     &\pdefin \;\bang\pTrans{A} \lto \pTrans{B} \\[2mm]
\pTrans{(\forall x A)} &\pdefin \forall x \pTrans{A}                \\[2mm]
\pTrans{(\exists x A)} &\pdefin \exists x \pTrans{A}.
\end{array}
}
If $A$ is provable in $\ILomega$ then $\pTrans{A}$ is provable in $\ILLomegar + \Pdis + \Pexi$, where
\begin{itemize}
\item[$\Pdis$] $\; \colon \;\; \bang(A \oplus B) \lto \,\bang A \,\oplus \,\bang B$
\item[$\Pexi$] $\; \colon \;\; \bang \exists x A \lto \exists x \bang A$.
\end{itemize}
\end{prop}
\proof First we show that given the principles $\Pdis$ and $\Pexi$, we have $\bang \lTrans{A}\lequiv ~\bang \pTrans{A}$. The
proof is done by induction on the complexity of the formula $A$.
Conjunction, implication and universal quantification follow easily
by induction hypothesis using that $\ILLomegar$ proves:
\eqleft{
\begin{array}{lcl}
\bang (A \awedge  B) & \lequiv & ~\bang A \,\cwedge \, \bang B \\[2mm]
\bang(\bang A\lto B) & \lequiv & ~ \bang(\bang A\lto \, \bang B) \\[2mm]
\bang \forall x A & \lequiv & ~ \bang \forall x \bang A
\end{array}
}
respectively. Disjunction and existential quantification are studied below:
\eqleft{
\begin{array}{lcl}
\bang \lTrans{(A\vee B)} & \equiv & \bang (\bang \lTrans{A}\oplus ~ \bang \lTrans{B})
   \lequiv \bang \lTrans{A} \oplus ~ \bang \lTrans{B} \\[2mm]
   & \stackrel{(\textup{IH})}{\lequiv} & \bang \pTrans{A}\oplus ~ \bang \pTrans{B}
   \stackrel{(\Pdis)}{\lequiv} \bang (\pTrans{A}\oplus \pTrans{B})\equiv ~\bang \pTrans{(A\vee B)}
\end{array}
}
and $\bang \lTrans{(\exists x A)}\equiv ~ \bang \exists x \bang
\lTrans{A}\lequiv \exists x  \bang
\lTrans{A}\stackrel{(\textup{IH})}{\lequiv}\exists x  \bang
\pTrans{A}\stackrel{(\Pexi)}{\lequiv} ~\bang \exists x
\pTrans{A}\equiv ~\bang \pTrans{(\exists x A)}$. Applying Proposition \ref{g-trans}, we know that from $\ILomega
\proves A$ we have $\ILLomegar \proves \lTrans{A}$. So, $\ILLomegar
\proves \, \bang \lTrans{A}$ and hence $\ILLomegar +
\Pdis + \Pexi \proves \, \bang \lTrans{A}$. Using the
equivalence proved before we have $\ILLomegar + \Pdis +
\Pexi \proves \, \bang\pTrans{A}$. In particular, we conclude
$\ILLomegar
+ \Pdis + \Pexi \proves \pTrans{A}$. \qed

The reason we are freely allowed to assume the principles $\Pdis$ and $\Pexi$ is that they are interpretable in all choices of
interpretations we consider. Let us argue that $\Pdis$ and $\Pexi$ are interpretable, by showing that the interpretation of the premise implies that of the conclusion (hence the identity and projection functions can be taken as realisers for the implication). For the three choices of $\ubq{x}{a}{A}$ we have considered one can show that
\begin{itemize}
   \item[] $\ubq{x}{a}{(A(x)  \awedge  B)} \lto(\ubq{x}{a}{A(x)}  \awedge  B)$ and
   \item[] $\ubq{x}{a}{(B\lto A(x))} \lto (B \lto \ubq{x}{a}{A(x)})$
\end{itemize}
when the variable $x$ does not occur free in $B$. Also, $\bang (\pcond{b}{A}{B})\lto \,\pcond{b}{\bang A}{\bang B}$. Therefore, we have that
\eqleft{
\begin{array}{lcl}
\uInter{\bang (A \oplus B)}{\pvec x, \pvec v, b}{\pvec a,\pvec c}
 & \equiv & ~\bang \ubq{\pvec y}{\pvec a}{\ubq{\pvec w}{\pvec c}{(\pcond{b}{\uInter{A}{\pvec x}{\pvec
y}}{\uInter{B}{\pvec v}{\pvec w}})}} \\[2mm]
 & \lto & ~\bang (\pcond{b}{\ubq{\pvec y}{\pvec a}{\uInter{A}{\pvec x}{\pvec y}}}{\ubq{\pvec w}{\pvec c}{\uInter{B}{\pvec v}{\pvec w}}}) \\[2mm]
 & \lto & ~\pcond{b}{\bang \ubq{\pvec y}{\pvec a}{\uInter{A}{\pvec x}{\pvec y}}}{\bang \ubq{\pvec w}{\pvec c}{\uInter{B}{\pvec v}{\pvec w}}}
 \equiv \uInter{\bang A~\oplus ~\bang B}{\pvec x, \pvec v, b}{\pvec a,\pvec c}.
\end{array}
}
Similarly, $\uInter{\bang \exists z A}{\pvec x, \pvec z}{\pvec a}\equiv ~\bang \ubq{\pvec y}{\pvec a}{\uInter{\exists z A}{\pvec x, \pvec z}{\pvec y}}\equiv ~\bang \ubq{\pvec y}{\pvec a}{\uInter{A}{\pvec x}{\pvec y}} \equiv \uInter{\bang A}{\pvec x}{\pvec a}\equiv \uInter{\exists z \bang A}{\pvec x, \pvec z}{\pvec a}$. Therefore, we can make use of the principles $\Pdis$ and $\Pexi$ to simplify the embeddings
of intuitionistic logic into (this extension of) linear logic, since the interpretation of linear
logic will interpret these principles taking us back to standard linear logic (without $\Pdis$ and $\Pexi$). This is illustrated in the following diagram, where $\ILLomega_{\textup P}$ abbreviates $\ILLomegar + \Pdis + \Pexi$ and $\ILomega_{{\sf ef}}$ abbreviates $\ILomega$ without disjunctions and existential quantifications:
\begin{center}
   \setlength{\unitlength}{10mm}
   \begin{picture}(6.0,3.0)
       \thicklines
       \put(0.3,2.5){$\ILLomega_{\textup P}$}
       \put(-0.2,1.2){$\pTrans{(\cdot)}$}
       \put(1.1,2.65){\vector(1,0){3.3}}
       \put(2.4,2.0){$\uInter{\cdot}{}{}$}
       \put(4.5,2.5){$\ILLomegab$}
       \put(4.95,1.2){$\pTrans{(\cdot)} = \lTrans{(\cdot)}$}
       \put(0.65,0.55){\vector(0,1){1.8}}
       \put(4.65,0.55){\vector(0,1){1.8}}
       \put(0.5,0.0){$\ILomega$}
       \put(1.2,0.15){\vector(1,0){3.2}}
       \put(1.8,0.4){Interpretation}
       \put(4.5,0.0){$\ILomega_{{\sf ef}}$}
   \end{picture}
\end{center}
The equality on the rightmost upward arrow represents the fact that all our interpretations transform proofs in $\ILomega$ into proofs in $\ILomega_{{\sf ef}}$, where the two translations $\lTrans{(\cdot)}$ and $\pTrans{(\cdot)}$ coincide.

\subsection{Modified realizability}

Kreisel's modified realizability associates with each formula $A$ of intuitionistic logic a new formula ``$\pvec x \mr A$", where $\pvec x$ is a sequence of fresh variables not present in $A$ (see \cite{Troelstra(73)} for the formal definition). We are going to prove that this form of realizability once translated to the linear logic context via $\bTrans{(\cdot)}$ corresponds (according to Theorem \ref{mr-rel} below) to the interpretation of $\ILLomegar$ with $\uInter{\bang
A}{\pvec x}{} \pdefin  ~\bang \forall \pvec y \uInter{A}{\pvec x}{\pvec y}$. First an auxiliary result:

\begin{lem}\label{circlebang}
$\uInter{\bTrans{A}}{\pvec x}{}\lequiv ~\bang \uInter{\bTrans{A}}{\pvec x}{}$.
\end{lem}
\proof Note that, because of the way we interpret $\bang A$, it can be checked by induction on $A$ that the interpretation of $\bTrans{A}$ has an empty tuple of challenge variables, i.e. we obtain a formula of the form $\uInter{\bTrans{A}}{\pvec x}{}$. To verify the lemma, it is enough to prove that $\uInter{\bTrans{A}}{\pvec x}{}\lequiv ~\bang A'$, for some formula $A'$, since assuming this we have $\bang \uInter{\bTrans{A}}{\pvec x}{}\lequiv ~\bang\bang A'\lequiv ~ \bang A'\lequiv \uInter{\bTrans{A}}{\pvec x}{}$. The proof is done by induction on the complexity of the formula $A$. We just sketch the cases of conjunction and disjunction, the other cases being immediate.
\eqleft{
\begin{array}{lcl}
\uInter{\bTrans{(A\wedge B)}}{\pvec x, \pvec y}{} & \equiv & \uInter{\bTrans{A}\cwedge \bTrans{B}}{\pvec x, \pvec y}{} \\[2mm]
	& \equiv & \uInter{\bTrans{A}}{\pvec x}{}\cwedge \uInter{\bTrans{B}}{\pvec y}{} \\[2mm]
	& \stackrel{(\textup{IH})}{\lequiv} & ~\bang A'\cwedge ~\bang B'\lequiv ~\bang (A' \awedge B'). \\[4mm]
\uInter{\bTrans{(A\vee B)}}{\pvec x, \pvec y,z}{}\equiv \uInter{\bTrans{A}\oplus \bTrans{B}}{\pvec x, \pvec y,z}{}
	& \equiv & \pcond{z}{\uInter{\bTrans{A}}{\pvec x}}{\uInter{\bTrans{B}}{\pvec y}{}} \\[2mm]
	& \stackrel{(\textup{IH})}{\lequiv} & \pcond{z}{~\bang A'}{\bang B'} \\[2mm]
	& \stackrel{(\textup{L\ref{useful}}(v))}{\lequiv} & ~\bang (\pcond{z}{\bang A'}{\bang B'}).
\end{array}}
That other cases are treated similarly. \qed

\begin{thm} \label{mr-rel} $\uInter{\bTrans{A}}{\pvec x}{}\lequiv \bTrans{(\pvec x \mr A)}$.
\end{thm}
\proof The proof is done by induction on the complexity of the formula $A$. If $A$ is an atomic formula, the result is trivial. Consider the case of conjunction:
\eqleft{
\begin{array}{lcl}
\uInter{\bTrans{(A\wedge B)}}{\pvec x, \pvec y}{} & \equiv & \uInter{\bTrans{A}\cwedge \bTrans{B}}{\pvec x, \pvec y}{}  \equiv \uInter{\bTrans{A}}{\pvec x}{}\cwedge \uInter{\bTrans{B}}{\pvec y}{} \\[2mm]
  & \stackrel{(\textup{IH})}{\lequiv} & \bTrans{(\pvec x \mr A)}\cwedge \bTrans{(\pvec y \mr B)} \\[2mm]
  & \equiv & \bTrans{(\pvec x \mr A \wedge \pvec y \mr B)}
  \equiv \bTrans{(\pvec x, \pvec y \mr A\wedge B)}.
\end{array}
}
The universal and existential quantifications also follow immediately using the induction hypothesis, and the way we define the translation and the interpretations. Implication is treated as
\eqleft{
\begin{array}{lcl}
\uInter{\bTrans{(A\to B)}}{\pvec g}{}
 & \equiv & \uInter{\bang (\bTrans{A}\lto \bTrans{B})}{\pvec g}{}\equiv~ \bang \forall x \uInter{\bTrans{A}\lto \bTrans{B}}{\pvec g}{\pvec x} \\[2mm]
 & \equiv & ~\bang \forall x (\uInter{\bTrans{A}}{\pvec x}{}\lto \uInter{\bTrans{B}}{\pvec g \pvec x}{}) \\[2mm]
 &  \stackrel{(\textup{IH})}{\lequiv}& ~\bang \forall x(\bTrans{(\pvec x \mr A)}\lto \bTrans{(\pvec g\pvec x \mr B)}) \\[2mm]
 & \lequiv & ~\bang \forall x\bang(\bTrans{(\pvec x \mr A)}\lto \bTrans{(\pvec g\pvec x \mr B)}) \\[2mm]
 & \equiv & \bTrans{(\forall x (\pvec x \mr A \to \pvec g \pvec x \mr B))}\equiv \bTrans{(\pvec g \mr (A\to B))}
\end{array}
}
whereas disjunction uses the auxiliary result above:
\eqleft{
\begin{array}{lcl}
\uInter{\bTrans{(A\vee B)}}{\pvec x, \pvec y, z}{}
 & \stackrel{(\textup{L}\ref{circlebang})}{\lequiv} & \bang \uInter{\bTrans{(A\vee B)}}{\pvec x, \pvec y, z}{} \equiv ~\bang \uInter{\bTrans{A} \oplus \bTrans{B}}{\pvec x, \pvec y, z}{} \equiv ~\bang (\pcond{z}{\uInter{\bTrans{A}}{\pvec x}{}}{\uInter{\bTrans{B}}{\pvec y}{}})\\[2mm]
 & \stackrel{(\textup{IH})}{\lequiv} & \bang ( (\bang (z=\true) \lto \bTrans{(\pvec x \mr A)}) \awedge (\bang (z=\false) \lto \bTrans{(\pvec y \mr B)})) \\[2mm]
 & \lequiv & \bang (\bang (z=\true) \lto  \bTrans{(\pvec x \mr A)}) \, \cwedge \, \bang (\bang (z=\false) \lto \bTrans{(\pvec y \mr B)}) \\[2mm]
 & \equiv & \bTrans{((z=\true \to \pvec x \mr A)\wedge (z=\false \to \pvec y \mr B))} \\[2mm]
 & \equiv & \bTrans{(\pvec x, \pvec y, z \mr A\vee B)}.
\end{array}
}
That concludes the proof. \qed

\subsection{G\"{o}del's Dialectica interpretation} \label{DialecticaSub}

Recall that G\"odel's Dialectica interpretation first associates with each formula $A$ a quantifier-free formula $\dInter{A}{\pvec x}{\pvec y}$ inductively. Then, each formula $A$ is interpreted as the new formula $\exists \pvec x \forall \pvec y \dInter{A}{\pvec x}{\pvec y}$ (see \cite{Avigad(98)}, section 2.3). The next result shows that the Dialectica interpretation corresponds to the $\ILLomegar$ interpretation where $\uInter{\bang A}{\pvec x}{\pvec y} \pdefin  ~\bang \uInter{A}{\pvec x}{\pvec y}$, via the simplified embedding $\pTrans{(\cdot)}$ (cf. Proposition \ref{simp-emb}).

\begin{thm}
$\uInter{\pTrans{A}}{\pvec x}{\pvec y}\lequiv \pTrans{(\dInter{A}{\pvec x}{\pvec y})}$.
\end{thm}
\proof The proof is again an easy induction on the complexity of the formula $A$. The atomic formulas are checked trivially and the other formulas follow immediately by induction hypothesis using the definitions of the $\pTrans{(\cdot)}$-translation and the interpretations. We illustrate with two cases: conjunction
\eqleft{
\begin{array}{lcl}
\uInter{\pTrans{(A \wedge B)}}{\pvec x, \pvec v}{\pvec y, \pvec w}
 & \equiv & \uInter{\pTrans{A} \awedge \pTrans{B}}{\pvec x,\pvec v}{\pvec y,\pvec w}
 \equiv \uInter{\pTrans{A}}{\pvec x}{\pvec y} \awedge  \uInter{\pTrans{B}}{\pvec v}{\pvec w} \\[2mm]
 & \stackrel{(\textup{IH})}{\lequiv} & \pTrans{(\dInter{A}{\pvec x}{\pvec y})} \awedge \pTrans{(\dInter{B}{\pvec v}{\pvec w})} \\[2mm]
 & \equiv & \pTrans{(\dInter{A}{\pvec x}{\pvec y}\wedge \dInter{B}{\pvec v}{\pvec w})}
  \equiv \pTrans{(\dInter{(A\wedge B)}{\pvec x, \pvec v}{\pvec y, \pvec w})}
\end{array}
}
and disjunction
\eqleft{
\begin{array}{lcl}
\uInter{\pTrans{(A\vee B)}}{\pvec x, \pvec v, z}{\pvec y, \pvec w}
 & \equiv & \uInter{\pTrans{A}\oplus \pTrans{B}}{\pvec x, \pvec v, z}{\pvec y, \pvec w}
 \equiv \pcond{z}{\uInter{\pTrans{A}}{\pvec x}{\pvec y}}{\uInter{\pTrans{B}}{\pvec v}{\pvec w}} \\[2mm]
 & \equiv & (\bang (z=\true) \lto \uInter{\pTrans{A}}{\pvec x}{\pvec y}) \awedge (\bang (z=\false)\lto \uInter{\pTrans{B}}{\pvec v}{\pvec w}) \\[2mm]
 & \stackrel{(\textup{IH})}{\lequiv} & (\bang (z=\true) \lto \pTrans{(\dInter{A}{\pvec x}{\pvec y})}) \awedge (\bang (z=\false)\lto \pTrans{(\dInter{B}{\pvec v}{\pvec w})}) \\[2mm]
 & \equiv & \pTrans{(z=\true \to \dInter{A}{\pvec x}{\pvec y})} \awedge  \pTrans{(z=\false \to \dInter{B}{\pvec v}{\pvec w})} \\[2mm]
 & \equiv & \pTrans{((z=\true \to \dInter{A}{\pvec x}{\pvec y})\wedge (z=\false \to \dInter{B}{\pvec v}{\pvec w}))} \\[2mm]
 & \equiv & \pTrans{(\dInter{(A\vee B)}{\pvec x, \pvec v, z}{\pvec y, \pvec w})}.
\end{array}
}
The other cases are treated similarly. \qed

Note that although $\pTrans{(\cdot)}$ translates formulas from $\ILomega$ into $\ILLomegar + \Pdis + \Pexi$, since these two principles are interpretable the verifying system is still $\ILLomega_b$.

\subsection{Diller-Nahm interpretation}

The Diller-Nahm interpretation differs from G\"odel's Dialectica interpretation since it allows finite sets to witness the negative content of an implication. Formally, the Diller-Nahm interpretation can be defined inductively as
\eqleft{
\begin{array}{lcl}
\dnInter{(\Aat)}{}{} & \pdefin & \Aat \\[2mm]
\dnInter{(A \wedge B)}{\pvec x, \pvec v}{\pvec y, \pvec w} & \pdefin & \dnInter{A}{\pvec x}{\pvec y}\wedge \dnInter{B}{\pvec v}{\pvec w} \\[2mm]
\dnInter{(A \vee B)}{\pvec x, \pvec v, z}{\pvec y, \pvec w} & \pdefin & (z=\true \to\dnInter{A}{\pvec x}{\pvec y})\wedge (z=\false \to \dnInter{B}{\pvec v}{\pvec w}) \\[2mm]
\dnInter{(A \to B)}{\pvec f, \pvec g}{\pvec x, \pvec w} & \pdefin & \forall y\in \pvec f \pvec x \pvec w \dnInter{A}{\pvec x}{\pvec y}\to \dnInter{B}{\pvec g\pvec x}{\pvec w} \\[2mm]
\dnInter{(\forall z A)}{\pvec f}{\pvec y, z} & \pdefin & \dnInter{A}{\pvec f z}{\pvec y} \\[2mm]
\dnInter{(\exists z A)}{\pvec x, z}{\pvec y} & \pdefin & \dnInter{A}{\pvec x}{\pvec y}. \\[2mm]
\end{array}}

Next we show that the Diller-Nahm interpretation of $\ILomega$ corresponds to the interpretation of $\ILLomegar$ with $\uInter{\bang A}{\pvec x}{\pvec a}  \pdefin \, \bang \forall \pvec y \!\in\! \pvec a \, \uInter{A}{\pvec x}{\pvec y}$.

\begin{thm}
$\uInter{\pTrans{A}}{\pvec x}{\pvec y}\lequiv \pTrans{(\dnInter{A}{\pvec x}{\pvec y})}$.
\end{thm}
\proof The proof, by induction on the structure of $A$, is similar to the one concerning G\"{o}del's interpretation. The only case which needs attention is that of implication, which we analyse below.
\eqleft{
\begin{array}{lcl}
\uInter{\pTrans{(A\to B)}}{\pvec f,\pvec g}{\pvec x, \pvec w}
 & \equiv & \uInter{\bang \pTrans{A}\lto \pTrans{B}}{\pvec f,\pvec g}{\pvec x,\pvec w}
 \; \equiv \; \uInter{\bang\pTrans{A}}{\pvec x}{\pvec f\pvec x\pvec w} \lto \uInter{\pTrans{B}}{\pvec g \pvec x}{\pvec w} \\[2mm]
 & \equiv & ~\bang \forall \pvec y\in \pvec f\pvec x\pvec w\uInter{\pTrans{A}}{\pvec x}{\pvec y}\lto \uInter{\pTrans{B}}{\pvec g \pvec x}{\pvec w} \\[2mm]
 & \stackrel{(\textup{IH})}{\lequiv} & ~\bang \forall \pvec y\in \pvec f\pvec x\pvec w\pTrans{(\dnInter{A}{\pvec x}{\pvec y})}\lto\pTrans{(\dnInter{B}{\pvec g\pvec x}{\pvec w})} \\[2mm]
 & \equiv & ~\bang \pTrans{(\forall \pvec y\in \pvec f\pvec x\pvec w\dnInter{A}{\pvec x}{\pvec y})}\lto \pTrans{(\dnInter{B}{\pvec g\pvec x}{\pvec w})} \\[2mm]
 & \equiv & \pTrans{(\forall \pvec y \in \pvec f\pvec x\pvec w \dnInter{A}{\pvec x}{\pvec y}\to  \dnInter{B}{\pvec g\pvec x}{\pvec w})} \\[2mm]
 & \equiv & \pTrans{(\dnInter{(A\to B)}{\pvec f,\pvec g}{\pvec x,\pvec w})}.
\end{array}
}
Note that the $\pTrans{(\cdot)}$ translation of $\forall y\!\in\! a \, A$ is $\forall y \!\in\! a \, \pTrans{A}$, as we can see below:
\eqleft{
\begin{array}{lcl}
\pTrans{(\forall y \!\in\! a \, A)} & \equiv & \pTrans{(\forall y (y\!\in\! a \to A))} \\[2mm]
 & \equiv & \forall y (\bang \pTrans{(y\!\in\! a)}\lto \pTrans{A})
 \; \equiv \; \forall y (\bang (y\!\in\! a)\lto \pTrans{A}) \; \equiv \; \forall y\!\in\! a \, \pTrans{A}.
\end{array}
}
That concludes the proof. \qed

\section{The Categorical Approach}
\label{categorical}

The study developed in this paper (and in previous work of the second
author) is strongly
inspired by work of de Paiva and Hyland on
categorical models of linear logic using G\"{o}del's Dialectica
interpretation. In this section we try to explain and make more explicit the link
between our framework for unifying interpretations of ${\sf IL}$ via
interpretations of ${\sf ILL}$ and the categorical approach on
\cite{dePaiva(1989A),dePaiva(1989B),dePaiva(1991)} for modelling ${\sf ILL}$.
More precisely, in \cite{dePaiva(1989A)} one finds a
categorical version of the Dialectica interpretation and an endofunctor interpretation for the modality $\bang A$ that corresponds to the
Diller-Nahm interpretation. Our goal is to relate this approach with
the work in the previous sections.

Before presenting de Paiva's category $\DC$ that models ${\sf
ILL}$, for sake of intuition, let us informally sketch the correspondence
between our framework and hers through the following table.
\[
\begin{array}{l|ll|cl}
& & \mbox{Our framework} & & \mbox{de Paiva's framework} \\[1mm]
\hline
\mbox{Realizers in} & & \textup{$\mathcal{T}^{\omega}$ - finite types} \quad & & \textup{$\C$ - cartesian closed category} \\[2mm]
\mbox{Formulas} & & \uInter{A}{}{}\subseteq X\times Y  & & X \stackrel{\alpha}{\nleftarrow} Y ~\textup{(object of $\DC$)} \\[2mm]
\mbox{Sequents} & & A\vdash B & & A\stackrel{(f,F)}{\longrightarrow}B ~\textup{(morphism of $\DC$)} \\[2mm]
%
%
\mbox{Linear implication} \quad & & A\lto B & & [A,B]_{\DC}~\textup{or}~B^{A}
%
%
%
\end{array}
\]
First, we point out that in de Paiva's work the realisers of the functional interpretation are taken from a given (fixed) cartesian closed category $\C$. In our case, we work with the particular cartesian closed category of the functionals of finite type. Also, our interpretations are given syntactically, and hence, a formula $A$ is interpreted as another formula $\uInter{A}{\pvec x}{\pvec y}$, which can be thought of as a binary relation between $\pvec x$ and $\pvec y$. In de Paiva's work these relations are at the core of constructing a new category $\DC$ out of the given ccc $\C$.

Let us briefly describe how the category $\DC$ is defined and its associated constructions. Starting with $\C$, a finitely complete cartesian closed category with stable and disjoint coproducts, we can define the monoidal closed category $\DC$ as follows. An object of $\DC$ is a subobject of the product $U\times X$, thus a monomorphism $A\stackrel{\alpha}{\rightarrowtail}U\times X$ with $A$, $U$ and $X$
objects of $\C$ also denoted by $U\stackrel{\alpha}{\nleftarrow} X$. If we think of these objects as set-theoretic relations between $U$ and $X$, and considering $\alpha$ as the identity monic, we get that $A \subseteq U \times X$, precisely as in our framework.

A map between two such objects $A\stackrel{\alpha}{\rightarrowtail}U\times X$ and $B\stackrel{\beta}{\rightarrowtail}V\times Y$ consists of a pair of maps of $\C$, $(f,F)$, $f:U\to V$, $F:U\times Y \to X$ such that pulling back $A\stackrel{\alpha}{\rightarrowtail}U\times X$ along $U\times Y\stackrel{(\pi_1,F)}{\longrightarrow}U\times X$ and $B\stackrel{\beta}{\rightarrowtail}V\times Y$ along $U\times Y\stackrel{f\times Y}{\longrightarrow}V\times Y$ (see the diagram below), the first subobject $A'\stackrel{\alpha'}{\rightarrowtail}U\times Y$ is smaller than the second $B'\stackrel{\beta'}{\rightarrowtail}U\times Y$, i.e. there is a map $k:A'\to B'$ in $\C$ making the triangle in the diagram below commute:

\begin{diagram}
   &                 &  A'             & \rTo             & A            \\
   &     \ldTo^{k}            &  \dMap^{\alpha'} &                  & \dMap^{\alpha}\\
B' & \rMap^{\beta'}   &   U\times Y      & \rTo^{(\pi_1,F)} & U\times X   \\
\dTo&                 &  \dTo^{f\times Y}&                  &             \\
B   & \rMap^{\beta}    &  V\times Y       &                  &             \\
\end{diagram}

If we write the two relations in the short version $U\stackrel{\alpha}{\nleftarrow} X$ and $V\stackrel{\beta}{\nleftarrow} Y$ and $(-)^{-1}$ for the pullback functor, then a map in $\DC$ can be represented as the pair $(f,F)$ in the diagram below

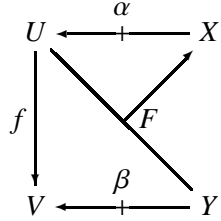
\begin{diagram}
U           & \lRel^{\alpha}  & X \\
\dTo^{f}     & \rdLine  \ruSmall \hspace{.7cm} F       &   \\
V            & \lRel^{\beta}   & Y  \\
\end{diagram}

\medskip\noindent satisfying the condition $(\pi_1,F)^{-1}(\alpha)\leq
(f\times Y)^{-1}(\beta)$.

The intuition in terms of set-theoretic relations is the following: there is a map $\alpha\stackrel{(f,F)}{\to}\beta$ in $\DC$ if and only if whenever $u\alpha F(u,y)$ then $f(u)\beta y$. In what follows we are going to say that two elements, $x$ and $y$ are related via $\alpha$ (i.e. $x\alpha y$) by $\alpha^{x}_{y}$. This way the comparison with our framework becomes easier. Using this notation, the condition above says that whenever $\alpha^{u}_{F(u,y)}$ then $\beta^{f(u)}_{y}$.

In the category $\DC$ we can also define the bifunctors $\cwedge$, $[-,-]_{\DC}$ and $\awedge$ and the operation $\oplus$ of weak-coproducts that can be read intuitively as
\[
\begin{array}{lll}
(\alpha\cwedge \beta)^{u,v}_{x,y} & \textup{iff}  & \alpha^{u}_{x}~ \textup{and} ~\beta^{v}_{y} \\[2mm]
[\alpha,\beta]^{F,f}_{v,z}\equiv(\beta^{\alpha})^{F,f}_{v,z} \quad & \textup{iff} & \alpha^{v}_{F(v,z)}\Rightarrow \beta^{f(v)}_{z} \\[2mm]
(\alpha \awedge \beta)^{u,v}_{w} & \textup{iff} \quad & \alpha^{u}_{w} ~\textup{or}~ \beta^{v}_{w} ~\textup{depending whether $w$ is in $X$ or $Y$}\\[2mm]
(\alpha\oplus \beta)^{w}_{f,g} & \textup{iff} & \alpha^{w}_{f(w)} ~\textup{or} ~ \beta^{w}_{g(w)} ~\textup{depending whether $w$ is in $U$ or $V$.}
\end{array}
\]
Apart from the relation $\alpha \oplus \beta$, our interpretation of the linear logic connectives (Definition \ref{inter})
coincides precisely with the definitions above. Let us examine in more detail the interpretation of $\oplus$, where our two approaches lead to different interpretations.

The main reason why we can have a simpler definition of
$\alpha \oplus \beta$ (with no need for the second player to play
higher order moves $f, g$) is because we always assume that each finite type is
inhabited by at least one element, while de Paiva's imposes no
similar restriction. More precisely, in our setting we have
\eqleft{\uInter{\alpha \oplus \beta}{}{}\subseteq (U\times V \times \mathbb{B})\times (X\times Y),}
with $\mathbb{B}$ for the set of boolean constants, while in de Paiva's setting,
considering set-theoretic relations,
\eqleft{(\alpha\oplus \beta)\subseteq (U+V)\times (X^{U}\times Y^{V}).}
If $U$ and $V$ are non-empty, then the two types $U\times V \times \mathbb{B}$ and $U + V$ are isomorphic. In case, however, one of $U$ or $V$ is empty then $U \times V \times \mathbb{B}$ is also empty, whereas $U+V$ can still be non-empty. In other words, in the most general case, when types can be empty, we must indeed work with the type $U + V$ rather than with $U \times V \times \mathbb{B}$. Let us see then, how the interpretation of $\alpha \oplus \beta$ works in the case when some of the move-sets of Eloise could be empty.

While in the first situation Eloise plays one element of $U$, one from $V$, and a boolean choosing which game is going to count, in the
second case Eloise plays an element of $U+V$. As we are going to see, in the latter case (with no extra assumptions) we need Abelard to play functions.
%
%
%
Consider the $\oplus\textup{L}$-rule, where from $\Gamma, A \vdash C$ and $\Gamma, B \vdash C$ we can conclude $\Gamma,A\oplus B \vdash C$ (for simplicity we shall omit the context $\Gamma$). In our framework, the proof of the two premises will provide realisers $F$ and $H$ such that the premises of the following rule are derivable:
\[
\begin{prooftree}
   \uInter{A}{u}{F u w}\lto \uInter{C}{f u}{w} \quad \quad  \uInter{B}{v}{H v w}\lto \uInter{C}{h v}{w}
   \justifies
   \uInter{A\oplus B}{u,v,b}{F u w, H v w}   \lto \uInter{C}{b(f u, h v)}{w}
\end{prooftree}.
\]
In order to realise $C$ in the conclusion of the rule we can make use of the boolean $b$ to choose between $f u$ and $h v$. Moreover, the ``negative" realiser for $A \oplus B$ is just a pair $\langle F u w, H v w \rangle$. Now, in de Paiva's more general setting, we have the same information about the realisers for the premises of the rule
\[
\begin{prooftree}
   \uInter{A}{u}{F u w}\lto \uInter{C}{f u}{w} \quad \quad  \uInter{B}{v}{H v w}\lto \uInter{C}{h v}{w}
   \justifies
   \uInter{A\oplus B}{a}{F' a w, H' a w}   \lto \uInter{C}{(a \in U)(f a)(h a)}{w}
\end{prooftree}.
\]

\medskip\noindent
but in the conclusion we no longer have a triple $\langle u, v, b \rangle$, but rather an element $a \in U + V$. Therefore, the functionals $F$ and $H$ are lifted to functionals $F'$ and $H'$ as
%
\medskip
\eqleft{F' a w : =
\left\{
\begin{array}{ll}
 \lambda u^U . F a w \quad & {\sf if} \; a \in U \\[2mm]
 \lambda u^U . F u w & {\sf if} \; a \in V
\end{array}
\right.
~ \quad \quad ~H' a w :=
\left\{
\begin{array}{ll}
 \lambda v^V . H v w \quad & {\sf if} \; a \in U \\[2mm]
 \lambda v^V . H a w & {\sf if} \; a \in V.
\end{array}
\right. }

\medskip\noindent
The extra arguments $u$ and $v$ are used in the cases when the parameter $a$ has the ``wrong" type to be used in either $F$ or $G$, and a standard value must be used.


For the rest of the section, let us analyse how the linear logic exponential $\bang A$ is interpreted in both approaches. As pointed by Blute and Scott in \cite{Blute(04)}, apropos natural and satisfying categorical models for the ${\sf LL}$ connectives ``\emph{unfortunately, the exponentials are less clear: the structure seems less canonical'}'.
In terms of monoidal categories the structure used to model $\bang A$ is that of comonads and comonoid objects.
In \cite{dePaiva(1989A)}, it is shown that if the category $\C$ has a free monoid structure with countable coproducts then the endofunctor $\bang$ can be defined on objects of $\DC$ as the pullback of $A^* \stackrel{\alpha^*}{\rightarrowtail}(U\times X)^*$ along $U\times X^*\stackrel{C_{(U, X)}}{\longrightarrow}(U\times X)^*$:

\begin{diagram}
\bang A             & \rTo             & A^*          &          & A  \\
\dMap^{\bang \alpha} &                  & \dMap^{\alpha^*} &         \lTo^{*} & \dMap_{\alpha} \\
U\times X^*      & \rTo^{C_{(U,X)}} & (U\times X)^*  &           & U\times X \\
\end{diagram}

\bigskip\noindent
Note that the functor $*:\C\to {\sf Mon} \C$ is left-adjoint to the
forgetful functor $U:{\sf Mon}\C \to \C$ (see \cite{dePaiva(1989A)} for more details).
Intuitively, the relation $\alpha^{u}_{x}$ is transformed into a new
relation $(\bang \alpha)^{u}_{\{x_1,\ldots ,x_n\}}$ which is
equivalent to $\forall x \!\in\! \{x_1,\ldots,x_n\} \, \alpha^{u}_{x}$.
The functor $\bang$ acts on morphisms in $\DC$ as $\bang
(f,F):\equiv (f,\bang F)$ where $\bang F:U\times Y^*\to X^*$ is the
composite of
\[ U\times Y^* \stackrel{C_{(U,Y)}}{\longrightarrow}(U\times Y)^* \stackrel{F^*}{\longrightarrow} X^*. \]
Since the functor $\bang :\DC\to \DC$ has a natural comonad $(\bang,
\epsilon, \delta)$ structure and $\bang A$ is a comonoid object in
$\DC$, $\bang$ models the linear logic exponential in the style of the Diller-Nahm variant of
the Dialectica interpretation, via finite sets.

In our approach, we have chosen to take a formal (syntactic) approach for the interpretation of $\bang A$. We identify
three conditions (A1--A3) which $\bang A$ needs to satisfy in order for the resulting interpretation to be sound. Our conditions are more general, and include as particular case the instance where $\bang$ is a comonad with comonoid objects. In particular, we are able to obtain interpretations of $\bang A$ that correspond to other well-known functional interpretation such as G\"odel's Dialectica interpretation and Kreisel's modified realizability.
A natural question, of course, arises: Do the Dialectica and modified realizability interpretations fit into the framework of de Paiva as well, and can they be seen as arising from other comonads with comonoidal structure?
In the first case (the Dialectica interpretation) the answer is yes, and de Paiva does have a few remarks about the Dialectica interpretation in her paper and in her thesis. More precisely, let $\bang : \DC \to \DC$ be the identity endofunctor. Intuitively $(\bang
\alpha)^{u}_{v}$ if and only if $\alpha^{u}_{v}$. It is immediate to check that
$(\bang,id,id)$ is a comonad, but in order for $\bang \alpha$ to be
a comonoid object in $\DC$ (not surprisingly) we need to require
decidability. More precisely, $\bang \alpha \to (\bang \alpha \,\cwedge\, \bang \alpha)$ is interpreted as $(\bang \alpha)^{x}_{y_0\cdot y_1} \to (\bang
\alpha)^{x}_{y_0} \cwedge (\bang \alpha)^{x}_{y_1}$ with
\eqleft{y_0\cdot y_1 := \left\{
\begin{array}{ll}
y_0 \quad & {\sf if} \; \neg \alpha^{x}_{y_0} \\[2mm]
y_1 & {\sf otherwise}.
\end{array}
\right.}
As for modified realizability, it is not clear to us at the moment whether it can also be shown to arise from a different monoid (other than the free monoid) using a generalisation of de Paiva's construction. We plan to consider this question in our future investigations.

\vspace*{.5cm}

\noindent {\bf Acknowledgements}. We would like to thank Jaime Gaspar for discussions related to the interpretations of intuitionistic linear logic. In particular, Lemma \ref{circlebang} was first observed by Gaspar, and also appears in \cite{GO(2010)}. Many thanks also to the anonymous referees for the detailed revision and comments that so much improved the final version of this paper.

\bibliographystyle{plain}


\begin{thebibliography}{10}

\bibitem{Avigad(98)}
J.~Avigad and S.~Feferman.
\newblock G\"odel's functional (``{D}ialectica") interpretation.
\newblock In S.~R. Buss, editor, {\em Handbook of proof theory}, volume 137 of
  {\em Studies in Logic and the Foundations of Mathematics}, pages 337--405.
  North Holland, Amsterdam, 1998.

\bibitem{Biering(2008)}
B.~Biering.
\newblock Cartesian closed dialectica categories.
\newblock {\em Annals of Pure and Applied Logic}, 156(2--3):290--307, 2008.

\bibitem{Blute(04)}
R.~Blute and P.~Scott.
\newblock Category theory for linear logicians.
\newblock In T.~Ehrhard, P.~Ruet, J-Y. Girard, and P.~Scott, editors, {\em
  Linear Logic in Computer Science}, pages 1--52. Cambridge University Press,
  2004.

\bibitem{Diller(74)}
J.~Diller and W.~Nahm.
\newblock Eine {V}ariant zur {D}ialectica-interpretation der {H}eyting
  {A}rithmetik endlicher {T}ypen.
\newblock {\em Arch. Math. Logik Grundlagenforsch}, 16:49--66, 1974.

\bibitem{GO(2010)}
J.~Gaspar and P.~Oliva.
\newblock Proof interpretations with truth.
\newblock {\em Mathematical Logic Quarterly}, 56(6):591--610, 2010.

\bibitem{Girard(87B)}
J.-Y. Girard.
\newblock Linear logic.
\newblock {\em Theoretical Computer Science}, 50(1):1--102, 1987.

\bibitem{Goedel(58)}
K.~G{\"o}del.
\newblock {\"U}ber eine bisher noch nicht ben{\"u}tzte {E}rweiterung des
  finiten {S}tandpunktes.
\newblock {\em Dialectica}, 12:280--287, 1958.

\bibitem{Hyland(02)}
J.~M.~E. Hyland.
\newblock Proof theory in the abstract.
\newblock {\em Annals of Pure and Applied Logic}, 114:43--78, 2002.

\bibitem{Kreisel(59)}
G.~Kreisel.
\newblock Interpretation of analysis by means of constructive functionals of
  finite types.
\newblock In A.~Heyting, editor, {\em Constructivity in Mathematics}, pages
  101--128. North Holland, Amsterdam, 1959.

\bibitem{Oliva(2007A)}
P.~Oliva.
\newblock Computational interpretations of classical linear logic.
\newblock In {\em Proceedings of WoLLIC'07, LNCS 4576}, pages 285--296.
  Springer, 2007.

\bibitem{Oliva(2007)}
P.~Oliva.
\newblock Modified realizability interpretation of classical linear logic.
\newblock In {\em Proc. of the Twenty Second Annual IEEE Symposium on Logic in
  Computer Science LICS'07}. IEEE Press, 2007.

\bibitem{Oliva(2008)}
P.~Oliva.
\newblock An analysis of {G}{\"o}del's dialectica interpretation via linear
  logic.
\newblock {\em dialectica}, 62(2):269--290, 2008.

\bibitem{Oliva(2009A)}
P.~Oliva.
\newblock Functional interpretations of linear and intuitionistic logic.
\newblock {\em Information and Computation}, 208(5):565 -- 577, 2010.

\bibitem{dePaiva(1989A)}
V.~C. V.~de Paiva.
\newblock The {D}ialectica categories.
\newblock In J.~W. Gray and A.~Scedrov, editors, {\em Proc. of Categories in
  Computer Science and Logic, Boulder, CO, 1987}, pages 47--62. Contemporary
  Mathematics, vol 92, American Mathematical Society, 1989.

\bibitem{dePaiva(1989B)}
V.~C. V.~de Paiva.
\newblock A {D}ialectica-like model of linear logic.
\newblock In D.~Pitt, D.~Rydeheard, P.~Dybjer, A.~Pitts, and A.~Poign\'e,
  editors, {\em Category Theory and Computer Science, Manchester, UK}, pages
  341--356. Springer-Verlag LNCS 389, 1989.

\bibitem{dePaiva(1991)}
V.~C. V.~de Paiva.
\newblock The {D}ialectica categories.
\newblock Technical Report 213, Computer Laboratory, University of Cambridge,
  Jan 1991.

\bibitem{Schellinx(1991)}
H.~Schellinx.
\newblock Some syntactical observations on linear logic.
\newblock {\em Journal of Logic and Computation}, 1(4):537--559, 1991.

\bibitem{Troelstra(73)}
A.~S. Troelstra.
\newblock {\em Metamathematical Investigation of Intuitionistic Arithmetic and
  Analysis}, volume 344 of {\em Lecture Notes in Mathematics}.
\newblock Springer, Berlin, 1973.

\end{thebibliography}

\end{document}